\documentstyle[emulateapj,apjfonts]{article}

\def\mkfigbox#1#2{
\centerline{
\hbox{ \epsfxsize=#1 \epsfbox{#2} \relax}
}
}

\def\Fig{{\footnotesize {\sc Fig.\/}~\,\thefigure. (continued) ---}}

\def\kms{km s$^{-1}$} 
\def\etal{{\it et al.}}

\def\Sec{${}^{\prime\prime}$\llap{.}}

\def\etal{{\it et~al.\/}}

\def\kms{{km~s$^{-1}$}}
\def\kpc-1{{kpc$^{-1}$}}
\def\Mpc-1{{Mpc$^{-1}$}}
\def\s-1{{sec$^{-1}$}}
\def\pdeg2{{deg$^{-2}$}}

\def\h0{{H$_0$}}
\def\q0{{$q_0$}}

\def\rms{{\it rms\/}}

\def\etal{{\it et al.\/}}
\def\kms{\hbox{$\rm km\,s^{-1}$}}

\def\ltsima{$\scriptscriptstyle \; \buildrel < \over \sim \;$}
\def\simlt{\lower.3ex\hbox{\ltsima}}

\def\gtsima{$\scriptscriptstyle \; \buildrel > \over \sim \;$}
\def\simgt{\lower.3ex\hbox{\gtsima}}

\def\about{\raise.3ex\hbox{$\scriptscriptstyle \sim $}}

\def\Sec{\hbox{${}^{\prime\prime}$\llap{.}}}

\def\sqr#1#2{{\vcenter{\vbox{\hrule height.#2pt
        \hbox{\vrule width.#2pt height#1pt \kern#1pt
        \vrule width.#2pt}
        \hrule height.#2pt}}}}
\def\square{{\mathchoice\sqr62\sqr62\sqr{4.2}1\sqr{3}1}\,}

\lefthead{Kelson \etal}
\slugcomment{Accepted for publicaton in the ApJ}

\begin{document}

\title{THE EVOLUTION OF EARLY-TYPE GALAXIES IN DISTANT CLUSTERS I.:
SURFACE PHOTOMETRY AND STRUCTURAL PARAMETERS FOR 53 GALAXIES
IN THE $z=0.33$ CLUSTER CL1358+62$^1$} 

\author{Daniel D.  Kelson\altaffilmark{2,3}, Garth D.
Illingworth\altaffilmark{3}, Pieter G. van Dokkum\altaffilmark{4,5}, and
Marijn Franx\altaffilmark{5}}

\altaffiltext{1}{Based on observations with the NASA/ESA {\it Hubble Space
Telescope}, obtained at the Space Telescope Science Institute, which is
operated by AURA, Inc., under NASA contract NAS 5--26555.}

\altaffiltext{2}{Department of Terrestrial Magnetism, Carnegie Institution
of Washington, 5241 Broad Branch Rd., NW, Washington, DC 20015}

\altaffiltext{3}{University of California Observatories / Lick Observatory,
Board of Studies in Astronomy and Astrophysics, University of California,
Santa Cruz, CA 95064}

\altaffiltext{4}{Kapteyn Astronomical Institute, P.O. Box 800, NL-9700 AV,
Groningen, The Netherlands}

\altaffiltext{5}{Leiden Observatory, P.O. Box 9513, NL-2300 RA, Leiden, The
Netherlands}

\begin{abstract}

Using wide-field, two-color HST imaging of the cluster CL1358+62
($z=0.33$), we derive structural parameters for a large,
magnitude-limited sample of cluster members. These structural
parameters are combined with accurate velocity dispersions in another
paper to investigate the fundamental plane in the cluster. We fit
integrated $r^{1/4}$-laws to the integrated surface brightness
profiles, and fit two-dimensional $r^{1/4}$-law model galaxies to the
images directly. A comparison of the results from the two methods
shows that the derived half-light radii, $r_e$, agree very well, with
an \rms\ scatter of only 13\%. The half-light radii range from
approximately 1 to 20 kpc, with a median of about 3 kpc
($H_0=65$\kms\Mpc-1, $q_0=0.1$).

We investigated the stability of $r_e$ by comparing the $r^{1/4}$-law
fits to the half-light radii derived using other commonly used surface
brightness profiles. In particular, we fit Sersic $r^{1/n}$-laws
(finding the best-fit $n$ ranging between $n=1$-$6$) and
superpositions of $r^{1/4}$-law bulges with exponential disks. The
half-light radii derived from the best-fit Sersic profiles varied with
respect to $r_e$ from the $r^{1/4}$-law fits by only 1\% in the
median, but with a standard deviation of more than 40\% in $r_e$. For
the bulge-plus-disk fits, the derived half-light radii were offset
from $r_e$ of the $r^{1/4}$-law fits by 10\% in the mean, also showing
a large standard deviation of more than 40\%. By comparing the fitted
half-light radii from the Sersic laws with those derived from the
bulge-plus-disk fitting, one also finds a large scatter of 30\%. Based
on these tests, we conclude that, in general, half-light radii are
only measured with a typical accuracy of 30-40\%.

While there are large uncertainties in half-light radii, these do not
impact the subsequent fundamental plane analysis because the
combination $r\langle I\rangle^{0.76}$, which enters the fundamental
plane, is extremely stable. The \rms\ scatter in $r\langle
I\rangle^{0.76}$ is less than 3\%, regardless of the form of the
profile fit to the galaxies.

By fitting bulge-plus-disk profiles, we find that the median bulge
fraction of the sample is 84\% and that the few E+A galaxies in this
sample have disks which make up about 10-35\% of their total light.
These results are consistent with the residuals from fitting
two-dimensional $r^{1/4}$-law models directly to the galaxies, in
which disk-like structures are revealed in about half of the sample.
Two of the three E+A galaxies show spiral arm structure.

\end{abstract}


\section{Introduction}

Galaxy scaling relations provide a unique opportunity to study the
evolution of stellar populations through intermediate redshifts
(\cite{franx93,vogt}, 1997). Since their discovery (\cite{fj76,tf77}),
such scaling laws have been used primarily to measure distances, with
a reliability seemingly limited only by measurement errors and the
homogeneity of the galaxies in question. The Hubble Space Telescope
(HST) and WFPC2 have enabled high $S/N$ imaging of intermediate
redshift galaxies with a resolution similar to ground-based imaging of
galaxies in the Coma cluster. Using the high spatial resolution and
$S/N$ of WFPC2, and efficient spectrographs at large aperture
telescopes like the Keck 10m, we can directly measure galaxy scaling
relations up to redshifts of unity in an effort to understand their
origins, and how galaxies evolve ({\it e.g.\/},
\cite{schade,vdfp98,simard99}).

Of the known scaling relations, the fundamental plane
(\cite{faber87,dd87}) has the lowest observed intrinsic scatter, only
\about 15\% in distance for a given galaxy (\cite{jfk93,kelson}). The
fundamental plane (FP) is an empirical correlation between effective
radius, $r_e$, central velocity dispersion, $\sigma$, and surface
brightness, $\langle I\rangle_e$, for early-type galaxies. In a study
of 11 nearby clusters J\o{}rgensen \etal\ (1996) found, in Gunn $g$,
\begin{equation}
r_e \propto \sigma^{1.16} \langle I\rangle_e^{-0.76},
\end{equation}
with a low ({\it rms\/}) scatter of $\pm 20\%$ dex in $r_e$. In Coma,
the scatter is even lower, $\pm 14\%$ in distance
(\cite{lucey,jfk93}). The fundamental plane of Virgo ellipticals also
has a very low internal scatter of $\pm 12\%$ in distance
(\cite{kelsonh0}). Using the virial theorem and assuming that
early-type galaxies are an homologous family, the low scatter in the
fundamental plane implies a scatter of only $\pm 20\%$ in $M/L$ ratio
for a given a galaxy mass.

The utility of the fundamental plane for probing the luminosity
evolution of early-type galaxies has been clearly demonstrated by van
Dokkum \& Franx (1996), Kelson \etal\ (1997), and van Dokkum \etal\
(1998b), who measured the evolution of the fundamental plane
zero-point out to $z=0.83$. The implied evolution of the $M/L_V$
ratios is consistent with a high redshift of formation for stars in
early-type galaxies in rich clusters. Projections of the fundamental
plane, such as the $r\hbox{-}L$ relation (\cite{korm}) and the
1-parameter $L\hbox{-}\sigma$ Faber-Jackson relation (\cite{fj76}),
have also been used by other authors to study early-type galaxy
evolution to high redshifts
(\cite{schade,ziegler,bender98,pahre98}).

Other studies of high redshift cluster galaxies have confirmed the
notion that early-type galaxies are homogeneous, and likely to be old
(\cite{schade,ellis97}). Furthermore, Dressler \etal\ (1997) showed
that there has been substantial evolution in the numbers of S0
galaxies between redshifts $z\sim 0.5$ and the present. However, many
such results derive from limited imaging of cluster cores. Recently,
van Dokkum \etal\ (1998a) used the color-magnitude relation of
early-type galaxies to show that the uniformity of age extends to
large radii for the ellipticals of CL1358+62, and that the most recent
star-formation in massive clusters has occurred in the outer parts, in
galaxies with disks. van Dokkum \etal\ showed that the stellar
populations of galaxies with disks are less homogeneous than pure
bulge systems, but only at cluster radii which are outside the
standard WFPC2 footprint (at the distances of intermediate redshift
clusters). Evolution in cluster galaxies may be best detected in
large, wide-field surveys of clusters: cluster membership, wide
spatial extent, and large samples inclusive of all morphological types
are all crucial for finding, and measuring the evolution in cluster
populations. Surveys which are specifically biased towards ellipticals
may miss much of the recent ($z<1$) evolution in cluster galaxies
because the samples are biased against those morphologies which have
undergone the most recent epochs of star-formation.

The first detailed study of membership in CL1358+62 was performed by
Fabricant \etal\ (1991), who measured a blue and active galaxy
fraction of \about 15\%. Complete membership to $R=21$ mag is
discussed by Fisher \etal\ (1998) and a small subsample was used by
Kelson \etal\ (1997) to find the fundamental plane zero-point for the
cluster. van Dokkum \etal\ (1998a) has made a complete analysis of the
Color-Magnitude relation of 194 confirmed cluster members within the
$8'\times 8'$ HST mosaic. The imaging consists of two colors (F606W
and F814W) in 12 CVZ pointings. This WFPC2 imaging of CL1358+62 is a
superb foundation for building a detailed history of cluster galaxies,
from the current epoch, through intermediate redshifts. Future work
will concentrate on mosaics of WFPC2 imaging in the clusters
MS2053--04 ($z=0.58$) and MS1054--03 ($z=0.83$).

Since the publication of Kelson \etal\ (1997), we have completed the
surface photometry for a high-resolution spectroscopic sample of
galaxies within the HST mosaic's field. This subset of the large
membership database will be used for analysis of the fundamental plane
and absorption line strength properties. The data are discussed in \S
\ref{data}. The photometric calibration, the transformation to
redshifted Johnson colors, and sources of systematic calibration
errors, are discussed \S \ref{calib}. The techniques used to fit
$r^{1/4}$-law profiles to the data are discussed at length in \S
\ref{deriv}. In \S \ref{psferr} we discuss such instrumental issues as
the under-sampling and point-spread function errors. Deviations from
the de Vaucouleurs profile are explored at some length in Sections
\ref{genprof} and \ref{scomp} using $r^{1/n}$-laws (\cite{sersic}) and
bulge-disk decomposition of the circularized profiles. Tables of the
structural parameters are given in \S \ref{adopt} and our results are
summarized in \S \ref{conclusions}.


\section{The Data}
\label{data}

The HST imaging consists of 12 contiguous pointings, making up a large
mosaic of the cluster environment, about $1.5h^{-1}$ Mpc on a side
(where $H_0=100h$ \kms\Mpc-1). Each pointing consists of three 1200 s
exposures in each of F606W and F814W. The data were preprocessed by
the standard pipeline procedures which are discussed elsewhere
(\cite{holtz}).

The analysis presented in this paper was performed on the images
processed by van Dokkum \etal\ (1998a), who cleaned the images of
cosmic rays, modeled and removed the complicated background, and
combined the individual exposures. The most demanding task was the
removal of the scattered earth light that resulted from the
observations being performed in the Continuous Viewing Zone (CVZ) (see
\cite{vdcm}). In some of these observations, the background was higher
by factors of two to three. This scattered light increases the counts
in the sky background, and the secondary mirror supports cast shadows
in the pattern of an "X," in which the background counts are typically
not elevated. The excess in the background, and contrast between the
shadowed and illuminated regions of the CCDs can vary greatly, and its
magnitude depends on the position of the telescope in its orbit when a
CVZ exposure is taken. Fortunately, in each set of three exposures, in
a given pointing and filter, at least one frame had minimal or no
scattered light, thereby allowing accurate modeling of the scattered
earth light. Large-order ($n\simgt 15$) bivariate polynomials were fit
to the background and subtracted. More details of this modeling are
given by van Dokkum \etal\ (1998a).

The selection of our fundamental plane sample is discussed in Kelson
\etal\ (1999a). Briefly, we randomly selected cluster members within
the field of view of the HST mosaic, to $R \le 21$ mag. The selection
was performed to efficiently construct multi-slit plates for the
Low-Resolution Imaging Spectrograph (\cite{okelris}). We used three
masks, with different position angles on the sky. The region of
maximum overlap is in the center of the cluster, and thus the FP
sample is concentrated towards the core of the cluster.

Morphology was not a factor in the selection of our sample. In the
random selection process, three E+A galaxies were included; these will
be compared with those cluster members that have normal, early-type
spectra. The sample is about $50\%$ complete for $R\le 20.5$ mag (see
\cite{fish} for details on the statistical completeness of the
original redshift catalog). Three galaxies fainter than $R=21$ mag
were added to test the quality of velocity dispersion measurements for
faint galaxies. These spectroscopic observations and their reductions
are detailed in Kelson \etal\ (1999a). In total, we have central
velocity dispersions for 55 galaxies in the cluster. Unfortunately,
two of the galaxies were imaged too close to the WFPC2 CCD edges to
obtain reliable structural parameters. Thus, there are a total of 53
cluster members in our determination of the fundamental plane in
CL1358+62.

Images of these 53 galaxies are shown in Figure \ref{images}. The
images clearly show a wide range of morphologies, from ellipticals to
early-type spirals. Optical morphologies have been taken from
Fabricant \etal\ (1999). These authors classified several hundred
galaxies in the HST mosaic according to morphological type, $T$. The
galaxies in our high-resolution spectroscopic sample have $T\in
\{-5,-4,-3,0,1,2,3\}$ (E, E/S0, S0, S0/a, Sa, Sab, Sb, respectively).
The galaxy morphologies are listed in Table \ref{params}, in which
there are 11 E, 6 E/S0, 14 S0, 13 S0/a, 6 Sa, 2 Sab galaxies and 1 Sb
galaxy. This distribution is quite similar to nearby massive clusters
({\it e.g.\/}, \cite{oemler}).


\section{Photometric Calibration}
\label{calib}

The aim of this work is to produce surface photometry for use in the
analysis of the fundamental plane in the cluster. To enable comparison
of the CL1358+62 fundamental plane with equivalent observations of
nearby galaxies, the calibrated HST photometry must be transformed to
a standard system. Such transformations to a photometric system of
redshifted Johnson $B$ and $V$ bandpasses will allow us to compare our
observations directly to observations of nearby galaxies. Precisely
these transformations and calibration steps have been used previously
by Kelson \etal\ (1997) and van Dokkum \etal\ (1998a). Each step has
its own uncertainties, which will be summarized at the end of the
section.

\subsection{Zero-points}

In this section, we describe the multiple components in the zero-point
calibration of the HST photometry in F606W and F814W. The starting
points of the calibration are the zero-points for F606W and F814W
derived by Holtzman \etal\ (1995) using photometry of stars (point
sources) with an aperture of $r=0\Sec 5$. Our measured surface
brightnesses have an effective aperture defined by the extent of our
adopted point-spread functions, in this case $r=1\Sec 5$. Thus, we
must correct our magnitudes from the $r=1\Sec 5$ aperture to an
effective aperture of $r=0\Sec 5$. Using the encircled energy curves
in the {\it HST Data Handbook\/}, we estimate the correction to be
$+0.067$ mag. The Holtzman \etal\ (1995) calibration is also valid for
data taken using a gain of $14 e^{-}$/ADU. Therefore we also
incorporate the gain ratios for the individual CCDs as given by the
handbook. Furthermore, the HST Distance Scale Key Project has measured
a difference of $0.050$ mag between long and short exposures of
identical fields (see, {\it e.g.\/}, \cite{hill,stetson}) such that
long integrations have more counts than would be expected by simply
scaling the shorter exposures. Thus, the WFPC2 zero-points, which were
derived from short integrations of star fields, are not appropriate
for long integrations, and need to be corrected by $+0.050$ mag.

In summary, the F606W and F814W magnitudes are defined by
\begin{eqnarray}
{\rm F606W} = -2.5\log (DN/s) + 21.419 + 2.5\log (GR_i) + 0.067 + 0.050\nl
{\rm F814W} = -2.5\log (DN/s) + 20.840 + 2.5\log (GR_i) + 0.067 + 0.050
\end{eqnarray}
where $GR_i$ represents the ratios of the high- to low-gain modes for
the four WFPC2 CCDs, respectively: $ 1.987,2.003,2.006,1.955$. The
total exposure time for each point was 3600 s per filter.

\subsection{Transformation to redshifted $V$, $(B-V)$}

\begin{figure}
\mkfigbox{5in}{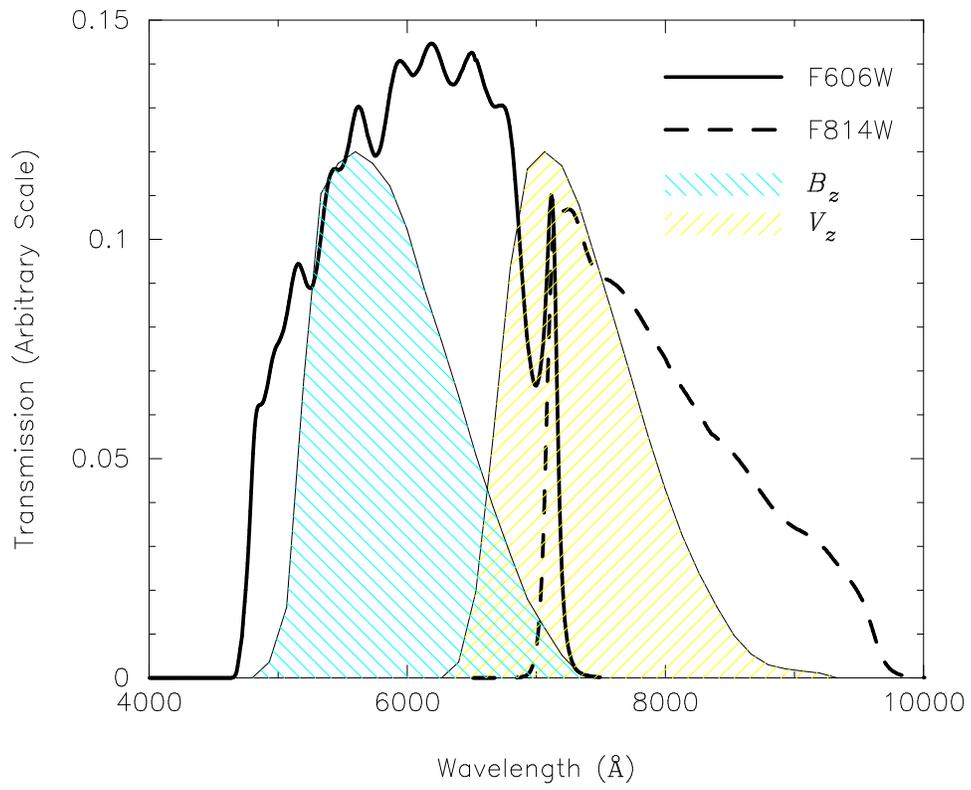}
\caption[Filter transmission curves]{
The transmission curves for the WFPC2 filters F606W and F814W are
superimposed with the Johnson $B_z$ and $V_z$ transmission curves
redshifted to $z=0.33$.
\label{filter}}
\end{figure}

In order to compare the surface brightnesses and $M/L$ ratios of
distant galaxies to equivalent observations made nearby, we must
transform the photometry to a consistent system. Towards this end, we
define $B_z$ and $V_z$ as the Johnson $B$ and $V$ filter bandpasses
redshifted to $z=0.33$, the observed frame of the galaxies. As one can
see in Figure \ref{filter}, the $B_z$ and $V_z$ bands fall at roughly
5500 \AA\ and 7000 \AA, overlapping substantially with the observed
F606W and F814W bandpasses. Without the color information, one would
be required to extrapolate to, {\it e.g.\/}, $V$ magnitude using $K$
corrections. However, we can derive an accurate transformation from
the WFPC2 system to the redshifted Johnson system by interpolation
using the color information.

As in van Dokkum \& Franx (1996), the flux in the redshifted filter
can be written as a function of the fluxes in the observed bandpasses:
\begin{equation}
F_{V_z} = F_{\rm F606W}^a \times F_{\rm F814W}^{(1-a)},
\end{equation}
where $a$ is a color term in this transformation (see below). In order
to relate the flux in a given bandpass to a standard magnitude, one
must use filter constants (see below). Solving for $V_z$ as a function
of the observed magnitudes and the filter constants, $c_X$, finds:
\begin{eqnarray}
&V_z = {\rm F814W} - c_{\rm F814W} + c_V +a\bigl[({\rm F606W-F814W}) -
&\nonumber\\
&(c_{\rm F606W}-c_{\rm F814W})\bigr] + 2.5\log(1+z)&
\end{eqnarray}
The additional factor of $2.5\log(1+z)$ compensates for the stretching in
wavelength of the redshifted bandpass. The $B_z$ and $V_z$ magnitudes are
redshifted measures of the fluxes, not restframe quantities. Removal of the
$(1+z)^4$ surface brightness dimming is required to convert these data to
magnitudes in the restframe of the galaxies.

The filter constants, $c_X$, are used to convert the $V_z$, F814W, and
F555W magnitudes to a system of flux units in a given filter. This
system can be thought of as similar to an AB system. These constants
are defined by $c_X=X-X_{AB}$, where $X$ is the magnitude, through
filter $X$, in the standard system. One can calculate the $AB$
magnitudes of Vega in the different filters, and, using the reported
magnitudes in the literature for Vega (\cite{holtz}), define the
transformation from the standard (Johnson) system to an $AB$ system,
based on absolute flux units. We used the Vega spectrum from Bohlin
(1996, \cite{hayes}), the filter transmission curves for the WFPC2
filters, and the Johnson filter curves (\cite{bessell}), as well as
the WFPC2 CCD response curve to determine the $AB$ magnitudes of Vega
in the various filters. These $AB$ magnitudes are zero-pointed to
Vega, not the spectra of F dwarfs ({\it i.e.\/} AB79, \cite{oke}; see
\cite{frei}, or \cite{fuk95,fuk96}).

Using the SEDs of E/S0 galaxies (\cite{pence}), one integrates over
the observed and redshifted filter curves to solve for $a$, for a
given SED:
\begin{equation}
a = {({V_z-c_V}) - ({\rm F814W} - c_{\rm F814W})\over
({\rm F606W} - c_{\rm F606W}) - ({\rm F814W} - c_{\rm F814W})}
\end{equation}
to find
\begin{eqnarray}
&V_z = {\rm F814W} + 0.204\times({\rm F606W}-{\rm F814W}) + 0.652&\\
&(B_z-V_z) = 0.818\times({\rm F606W}-{\rm F814W}) - 0.128&
\end{eqnarray}
for galaxies with E/S0 spectra at $z=0.33$. Using SEDs typical of
spirals yield transformations which differ by only $\pm 0.01$ mag. As
noted above, these transformations are identical to those used by
Kelson \etal\ (1997) and van Dokkum \etal\ (1998a).

\subsection{Colors}

Colors in the WFPC2 filters are taken directly from van Dokkum \etal\
(1998a). They measured $({\rm F606W - F814W})$ colors within $r_e$ for each
of the cluster members within the HST mosaic; their color errors (random)
are on the order of $\pm 0.01$ mag. For the purposes of the photometric
calibration, the absolute color uncertainties should approximately equal
the quadrature sum of the two Holtzman \etal\ (1995) zero-point errors.
Other effects, such as the exposure time correction, the charge-transfer
effect, gain variations and aperture corrections, cancel since they are
equal contributors to the calibration of each bandpass. We comment more on
color uncertainties below.

\subsection{Galactic extinction}

Galactic extinction has been taken from Burstein \& Heiles (1982) and
used in conjunction with Cardelli, Clayton, \& Mathis (1989) to derive
the extinctions in the HST filters. The Burstein \& Heiles $E(B-V)$
color excess is $0.006$ mag. Using Cardelli \etal, we find $R_{\rm
F606W}=2.91$ and $R_{\rm F814W}=1.94$. Thus, $A_{\rm F606W}=0.02$ and
$A_{\rm F814W}=0.01$. Thus, we correct both the measured colors and
the redshifted $V$ surface brightnesses by $0.01$ mag. Because the
Galactic reddening is low in the direction of CL1358+62, reddening
uncertainties will contribute little to the following error budget.


\subsection{Uncertainties in the Redshifted Photometry}

Errors in the final calibration of the redshifted $V_z$-band surface
brightnesses can arise from three sources: (1) errors in the HST
zero-point calibration, (2) errors in the measured colors, and (3)
errors in the transformation from the observed to redshifted
bandpasses. We now attempt to summarize these uncertainties, and
estimate the total error in our calibrated $V_z$ magnitudes.

Each step in the calibration of HST photometry introduces
uncertainties. In the previous section, we showed that the
$V_z$-bandpass is very closely tied to F814W, with a moderate color
term. Thus, errors in the Holtzman \etal\ (1995) zero-point for F814W
propagates directly to systematic errors in the $V_z$ magnitudes. The
F814W zero-point uncertainty is listed by Holtzman \etal\ (1995) as
$\pm0.006$ mag. Their calibration, however, is explicitly valid for
the spectral energy distributions of stars, so the calibration has
larger uncertainties when applied to the SEDs of galaxies, probably on
the order of 1-2\%. Furthermore, the Holtzman \etal\ (1995) zero-point
is valid for $0\Sec 5$ radius aperture magnitudes. Our surface
brightnesses will have been measured using data deconvolved with a
point-spread function with a finite radius ($1\Sec 5$). Therefore, we
must correct our data from an effective aperture of $r=1\Sec 5$ to an
effective aperture of $r=0\Sec 5$. A reasonable estimate for the
uncertainty in this correction is $\pm 0.01$ mag. Another source of
error arises from uncertainties in the Extragalactic Distance Scale
Key Project's correction to the zero-point (see, {\it e.g.\/},
\cite{stetson98}); we estimate that it introduces an uncertainty on
the order of $\pm 0.025$ mag (half the effect).

The color term in the transformation to redshifted $V_z$ is small,
with the result that errors in $V_z$ due to uncertainties in the $\rm
(F606W-F814W)$ colors are diminished by nearly a factor of five. These
color errors include the uncertainty in the F606W zero-point, which we
assume to be on the order of $\simlt 0.05$ mag, compared to the
estimate by Holtzman \etal\ (1995) of $\pm 0.005$ mag (the F606W
zero-point was determined theoretically, and no additional calibration
since Holtzman \etal\ (1995) has been attempted). With regard to the
colors, any systematic variations in the CCD gain and charge transfer,
for example, cancel. As stated earlier, the error in the adopted
Galactic reddening is likely to be small, no greater than $\pm 0.01$
mag. Thus, the primary component of systematic error is the F606W
zero-point. If the $\rm (F606W-F814W)$ color error can be thought of
as the quadrature sum of the Holtzman \etal\ (1995) zero-point errors,
then the resulting error in $V_z$ from galaxy colors is still $\simlt
0.01$ mag. Color gradients within the galaxies themselves could be a
larger factor, but van Dokkum \etal\ (1998a) showed that, in the mean,
the colors change very little as one adopts different standard radii
for measurement of the colors.

Errors in the coefficients in the transformation come from several
sources. Uncertainties in the filter constants are estimated to be
$\pm 0.03$ mag in the final transformation. Variations in the
intrinsic galaxy SEDs lead to (slightly) different values for the
color and constant terms in the transformations themselves,
introducing an uncertainty of $\pm 0.01$ mag. Lastly, adoption of the
Hayes (1985) spectrum of Vega may also lead to errors if the WFPC2
photometric calibration sources were not strictly on the same
spectrophotometric system as the Hayes spectrum. Unfortunately, the
magnitude of this uncertainty is not well known, though we estimate it
to be no larger than $\pm 0.05$ mag based on the agreement of several
sources of filter constants with with Holtzman \etal\ (1995).

By adding these contributions in quadrature, we estimate a total error
of $\pm 0.06$ mag in $V_z$ and $\pm 0.05$ mag in $(B_z-V_z)$ color.
For the analysis of the fundamental plane, these uncertainties can be
treated as systematic errors.


\section{Fitting de Vaucouleurs Profiles}
\label{deriv}

The photometric parameters, $r_e$ and $\langle I\rangle_e$, were
obtained by fitting $r^{1/4}$-law profiles to the galaxies and two
distinct approaches were used to fit $r^{1/4}$-laws to the data. The
first approach involved restoration of the data (deconvolution) using
the CLEAN procedure (\cite{hogbom}), isophote fitting using GALPHOT
(\cite{jfk92}), the derivation of a circularized, integrated surface
brightness profile ({\it i.e\/}, a ``growth'' curve), and its fit with
an integrated $r^{1/4}$-law.

The second approach involved fitting two-dimensional galaxy models to
the images directly (cf. \cite{vdf96}). The algorithm involves
convolving two-dimensional $r^{1/4}$-laws with a PSF, iteratively
adjusting the parameters until the model provides a good fit to the
WFPC2 images. Nearby, contaminating objects can be handled elegantly,
since multiple objects can be fit simultaneously (see also Simard
1998). For both of these procedures we used artificial PSFs for each
galaxy. These PSFs were created using Tiny Tim 4.0b (\cite{krist}). We
estimate the effects of PSF errors in \S \ref{psferr}.


\subsection{Fitting Integrated Surface Brightness Profiles}
\label{growthfit}

In this section, we discuss the procedures used for fitting the
integrated profiles derived from the WFPC2 images. Before deriving the
surface photometry, the images must be deconvolved because the
structural parameters $r_e$ and $I_e$ can be seriously affected by the
WFPC2 PSF and its broad wings. We chose to use the CLEAN method of
deconvolution because it is fast, efficient, and simple
(\cite{hogbom}). The output images were convolved with a Gaussian
smoothing function to reduce the noise in the output image. The
residual maps typically contained $\simlt 1\%$ of the counts in the
original image but were nevertheless added to the output frames in
order to preserve flux. We fitted elliptical isophotes using the
program GALPHOT (\cite{jfk92}). The radii were spaced logarithmically,
with an inner isophote radius of $r=0.3$ pixels, extending out to
radii typically five times larger than $r_e$.

In deriving structural parameters for our galaxies, we duplicated the
procedures used on nearby galaxies, fitting the integrated profiles
rather than the profiles themselves.


\subsubsection{Fitting for $r_e$}

The flux in the CLEANing process has been restored with an effective
PSF, chosen to be a Gaussian with $\rm FWHM\equiv 1$ pixel. At any
given radius, one needs to correct the analytical $r^{1/4}$-law growth
curve for the convolution by the restoration PSF. The enclosed flux
within a given radius for an $r^{1/4}$-law profile is analytically
simple:
\begin{eqnarray}
F(r) &=& \pi r_e^2 I_e {8e^{7.676} \gamma(8,x)\over{7.676^{-8}}},\nl
&=& \pi r_e^2 \langle I\rangle_e {2\gamma(8,x)\over 7!},\nl
\nonumber x &=& 7.676 \biggl({r\over r_e}\biggr)^{1/4}
\label{anal4}
\end{eqnarray}
The least-squares fit can be linearized straightforwardly, since the
derivatives of Equation \ref{anal4}, with respect to $r_e$ and $I_e$ (or
$\langle I\rangle_e$), are also analytically simple.

For a given set of structural parameters  $(r_e, I_e)$, one can
readily compute how much interior flux should be redistributed outside
of a given radius in the growth curve. In this fashion, one adjusts
the growth curve due to the smearing by the restoration PSF (see, {\it
e.g.\/}, \cite{saglia93}). At each radius, one computes a set of
aperture corrections to correct the model to a ``seeing''-convolved
model (assuming circular symmetry). These corrections are determined
numerically for each set of structural parameters tested in the search
for $\chi^2$ minimum. The expression for the convolved model growth
curves is
\begin{equation}
G(r) = \int_0^\infty \int_0^\infty 2\pi \rho I(\rho) P(k) J_0(kr)\,
J_1(k\rho) dk d\rho
\label{seeing}
\end{equation}
where the restoration PSF is described by its transform, $P(k)$, and
$J_0$ and $J_1$ are zeroth and first order Bessel functions. For
$I(\rho)$, we use the fitted profile. We use the Levenberg-Marquardt
method to perform the $\chi^2$ minimization, fitting $G(R)$ to the
integrated surface brightness profiles.\footnote{In order to linearize
the fitting procedure, we continue to describe the growth curve's
derivatives by their analytical expressions, multiplied by the ratio
of $G(R)/F(R)$, where $G$ and $F$ are the convolved and unconvolved
values of the cumulative flux.} Once the set of structural parameters
has been found which minimizes $\chi^2$, we have $r_e$, $I_e$, and
$\langle I\rangle_e$ for the fitted model. Once the model has been
fit, the galaxy profile may not be perfectly fit by the model within
$r_e$, the radius within which $\langle I\rangle_e$ is defined.
Therefore, we use the observed $\langle I\rangle_e$, {\it i.e.\/}, the
enclosed flux within $r_e$ for the observed profile instead of the
enclosed flux for the model, and have corrected this surface brightness
for the smearing effects of the restoration PSF (see Equation
\ref{seeing}). Using either the observed or model $\langle I\rangle_e$
has an effect on $r_e\langle I\rangle_e^{0.76}$ of less than $0.1\%$
in the mean.

\begin{figure}
\mkfigbox{7in}{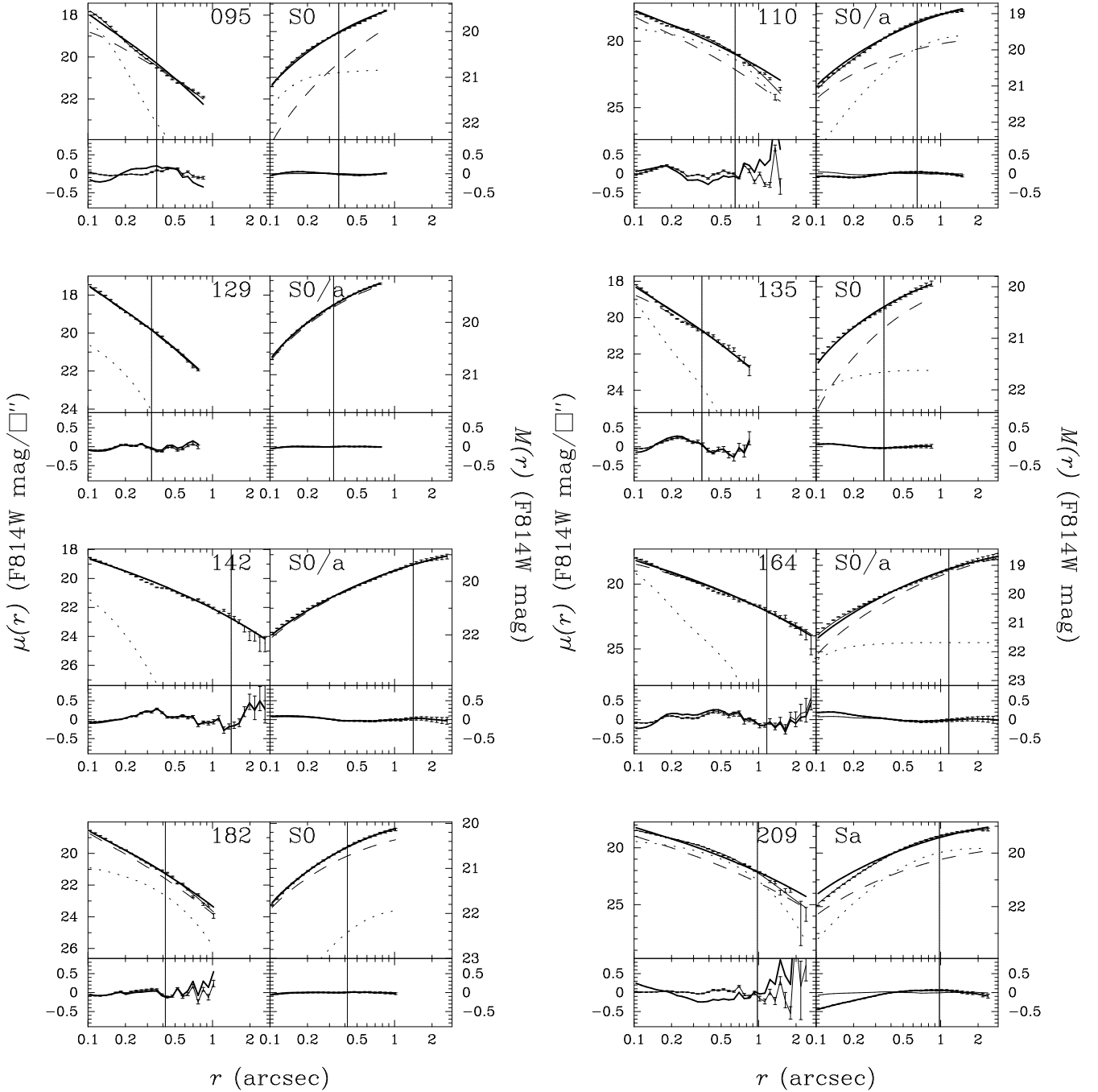}
\caption[Surface brightness and integrated surface brightness profiles]{
The F814W surface brightness profiles and integrated profiles from
$r=1$ pixel to \about$2r_e$. The thick solid lines show the fit of, and
residuals from, the fit of the $r^{1/4}$-law. The thin solid lines show
the fit of and residuals from the bulge-plus-disk superposition, with
the two components of the fit shown as dashed ($r^{1/4}$-law bulge)
and dotted (exponential disk) lines. The integrated curves were fit
between $r=1$ pixel and $2\,r_{1/2}$. Several galaxies are not well
fit by a pure de Vaucouleurs profile, though as noted in the text, the
effect of such deviations on the determinations of the fundamental
plane parameter, $r_e\langle I\rangle_e^{0.76}$, is quite small. The
profiles and growth curves are plotted over a range of $0\Sec 1 \le r
\le 5''$, except for galaxies \# 353, 371, and 375.
\label{profile}}
\end{figure}

In Figure \ref{profile}, we show the calibrated F814W surface
brightness profiles and integrated surface brightness profiles. The
fit by the integrated de Vaucouleurs profiles are shown by thick solid
lines in the right-hand panels. The implied $r^{1/4}$-law profiles are
shown in the left-hand panels. The residuals are shown in the lower
panels, using thick solid lines as well. In general, the quality of
the fit to the integrated profiles is good.


\subsubsection{Sensitivity to Fitting Range}

In the fitting process, one must select a range of radii over which to
fit the integrated surface brightness profiles. At small radii, the
profile shapes are affected by PSF errors which may have been
exacerbated by the (de)convolution procedures. Secondly, at large
radii the profiles themselves become noisy, and thus the growth curve
shapes can become suspect.

Three effects limit the choice of inner radius in the fit: (1) the
CLEAN procedure may not be perfect and noiseless; (2) the Tiny Tim
construction of the WFPC2 PSF may not be perfect; and (3) the
correction for the smearing by the restoration PSF may not be perfect.
These effects are exacerbated by the under-sampling of the data.

The outer radius was chosen in order to minimize the effects of
sky-subtraction errors. The shape of a galaxy's outer profile can be
greatly affected by errors in the sky background, even though fitting
of a parameterized growth curve is intended to be robust against such
uncertainties.

We find that the results are quite stable to the choice of inner and
outer fitting radii. We varied the outer fitting radius between $2.0$,
$2.5$ and $3\, r_e$, and the inner fitting radius between 0, 0.5, 1
pixel, and 2 pixels. (For $H_0=65$ \kms\Mpc-1 and $q_0=0.1$, one WFPC2
pixel corresponds to $0.48$ kpc, equivalent to $1''$ at the distance
of Coma.) For all combinations of the fitting range, mean and median
differences in $r_e$ are less than $\pm 1.5\%$. We found that $2\,
r_e$ was a suitable compromise for the outer radius in fitting the
growth curve. After investigating the uncertainties due to PSF errors
and the under-sampling (\S \ref{psferr}), we adopted an inner fitting
radius of $r=1$ pixel.

In the extreme case, changing the inner and outer fitting radii to
$r=0$ and $r=3\,r_e$, the mean systematic offset in $r_e$ is about
1\%, with an \rms\ scatter in the $r_e$ values of $9\%$. The
systematic offset in $r_e I_e^{0.76}$, the combination of the
structural parameters which appears in the fundamental plane, is
negligible, with an \rms\ scatter of 0.1\%. The stability of the
effective radii to the inner fitting boundary gives us confidence that
the procedure outlined above properly corrects for the convolution by
the restoration PSF.


\subsubsection{Sensitivity to Sky Subtraction}

One important source of error not addressed by varying the inner and
outer fitting radii was uncertainties in the determination of the sky
background. When the imaging data were processed in van Dokkum \etal\
(1998a), the background was modeled and subtracted using order
$n\simgt 15$ polynomials. However, for any given galaxy there may
still be residual background fluctuations that were not accurately
removed by the modeling. Thus, we followed the procedure in
J\o{}rgensen \etal\ (1992), where low-order power laws ($m=2,3$) are
fit to the outer parts of the profiles using an unweighted
least-squares routine. The sky value is estimated from the asymptotic
convergence of $I=I_0 r^{-m} + {\rm sky}$.

The \rms\ scatter in extrapolated sky values was 5$e^-$ (\about 24
mag/$\square ''$). By ignoring this fit for the local background,
the effect on $r_e$ is about 1\%, in the mean, with a standard
deviation of 3\%. No effect on the fundamental plane parameters was
found.

\medskip

In summary, the fundamental plane parameters derived from the fitting
of the integrated profiles, for all combinations of the inner and
outer fitting radii, are extremely robust against the choice of
fitting range. The insensitivity to the fitting range and to
uncertainties in the sky subtraction suggest to us that the
growth-curve parameters are well-determined.


\subsection{Fitting the Images Directly}
\label{r4fit}

In this section, we derive structural parameters by fitting
two-dimensional model galaxies directly to the WFPC2 images (see
\cite{vdf96} for a discussion of the technique used here, and other
authors for variants of the method {\it e.g.\/},
\cite{schade,simard}). Images of model galaxies, for a given set of
structural parameters, are convolved with the Tiny Tim PSFs and
compared directly to the data frames. The structural parameters are
adjusted until $\chi^2$, or some other measure of the goodness of fit,
is minimized. The local topology of $\chi^2$ is generally used to
estimate the uncertainties in the fit.

Procedures for fitting the images directly have the advantage that one
performs a convolution rather than deconvolution. The advantages of
such a process include the preservation of the image's noise
characteristics, and the fitting of multiple objects simultaneously.
The current disadvantage is that the 2D models are built using a
single, global ellipticity and position angle for each galaxy.
Nonetheless, as the apparent information content continues to shrink
with the decreasing angular sizes of high redshift galaxies, useful
structural parameters are still extractable via such modeling, so long
as the PSF is reasonably well known.

Such fitting methods assume a functional form to the galaxy profile,
such as an $r^{1/4}$-law, but can be expanded to include multiple
components ({\it e.g.\/}, $r^{1/4}$-law bulge plus exponential disk).
The parameters of the least-squares fit to an $r^{1/4}$-law are: (1)
$(x,y)$ position of the center; (2) $\epsilon$, the apparent
ellipticity; (3) position angle; (4) effective radius; (5) surface
brightness at the effective radius; and (6) the sky background.
Multiple-component ({\it e.g.\/}, bulge-plus-disk) fits require
additional parameters and complicated minimization algorithms, but
even such expanded profile shapes do not yet take into account
deviations from purely elliptical isophote shapes, or twisting of the
isophotes with radius.

Using the fitting procedure of van Dokkum \& Franx (1996), we derived
the $r^{1/4}$-law structural parameters for the galaxies. Given the
potential uncertainties in the processing of the sky background in the
CCD images, we performed the fitting first with the sky background
fixed at zero, and a second time while fitting for the local sky
background. When the sky background is fit, the $r_e$ changes by a
mean (median) of $+2\%$ $(+0.1\%)$, with a scatter of nearly $\pm
10\%$. The additional fitting of local sky has a negligible effect on
the fundamental plane parameter in the mean ($< 0.1\%$), with a
scatter of $0.5\%$. For some galaxies, $r_e$ could change by as much
as 30-40\%, but even for these extreme cases the fundamental plane
parameter only changes by $\simlt 2\%$.

In Figure \ref{residuals}, we show the galaxy images, after
subtraction with the best-fit 2D $r^{1/4}$-law models. Many disk-like
structures are clearly visible in these residuals and bulge-plus-disk
fits to the 1D profiles are discussed in \S \ref{scomp}.


\subsection{Comparison of Growth Curve and Image Modeling}
\label{growthvsdirect}

Ideally the two fitting procedures should yield similar, or even
identical structural parameters. In practice this will not be the case
for several reasons. For example, the two procedures have
intrinsically different weighting schemes in the fitting process.
Furthermore, galaxies do not have surface brightness profiles which
obey perfectly an $r^{1/4}$-law, or any other simply parameterized
function. While the internal tests listed above, such as varying the
fitting range of the integrated surface brightness profiles, or
allowing the image modeling to fit the sky background, help estimate
some of the uncertainties in the measurements, the methods themselves
suffer from their own systematics. Therefore, we now compare the
structural parameters derived from the two methods in order to help
estimate remaining systematic uncertainties.

The comparison of effective radii derived by the two algorithms is
surprisingly good, considering the range of apparent morphologies
covered by the sample. The median difference in effective radii
between the two algorithms is only $1\%$. The standard deviation is
$13\%$ in $r_e$, precisely the quadrature sum of the \rms\ scatter
derived in the individual internal tests. The median offset in
effective surface brightness is $-4\%$. Thus, there is a systematic
offset in the effective fundamental plane parameter of $1.5\%$, with a
standard deviation of $2.5\%$. Nominally, one would expect that the
error in the fundamental plane parameter should be smaller than the
error in the individual structural parameters, because of the
correlation between $r_e$ and surface brightness. However, the two
fitting methods weight the data differently and are therefore affected
differently by the under-sampling and PSF errors, which are addressed
in the next section. Nevertheless, these results indicate that the
errors in the fundamental plane parameters are small.


\section{Remaining Instrumental Effects}
\label{psferr}

While the above tests show that the fundamental plane parameters
appear well constrained, there are potential sources of error in
common to both methods. The two most important effects are (1) the
under-sampling; and (2) the use of model point-spread functions. Even
those these effects are common to both methods, the data are weighted
differently in the $\chi^2$ calculation. As a result, the two fitting
methods are not affected identically.

\subsection{The Effects of Under-sampling}

In order to test the effects of under-sampling, and the dependence of
the PSF on sub-pixel positioning, we generated a $10\times $
subsampled Tiny Tim PSF. We created a set of noiseless subsampled
$r^{1/4}$-law galaxy images using the subsampled PSFs. We then created
two sets of WF CCD sampled versions of these model galaxies. The first
set were simply block-averaged to the nominal WF CCD sampling and
convolved with a charge-smearing kernel (\cite{k97}). Thus, the galaxy
centroids were located at the precise corner of a WF CCD pixel. The
second set were created by shifting the subsampled model galaxy images
by $0\Sec 05$ (half a WF CCD pixel), then block-averaged, and
convolved with the charge-smearing kernel. In each case, the
block-averaged subsampled PSFs are equivalent to the (under-)sampled,
real ones. Thus, we had two sets of model galaxy images which differed
in position by half a WF CCD pixel. We then used the nominally sampled
Tiny Tim PSF, centered at the integer pixel location, to derive
structural parameters for these galaxies.

The model images were analyzed in the same manner as the CL1358+62
galaxies, as described in \S \ref{growthfit} and \S \ref{r4fit}. The
systematic errors in $r_e$ varied with galaxy size, as did the errors
in $r_e I_e^{0.76}$. For $r_e=4$ pixels, the error in the fundamental
plane parameter was only $-2\%$. For $r_e=2$ pixels, the error is
$-5\%$. For these tests, we restricted the growth curve fitting to
radii $r>1 $ pixel. By fitting to the very center, the systematic
error in $r_e I_e^{0.76}$ was nearly $-10\%$ for galaxies as small as
$r_e=2$ pixels. Even by directly fitting the images with the
two-dimensional models, the derived structural parameters are
susceptible to errors due to the under-sampling. For galaxies with
$r_e = 1$ pixel, the systematic errors due to under-sampling can be as
large as \about $50\%$ in $r_e$, and $\simlt 10\%$ in $r_e
I_e^{0.76}$. Fortunately, the smallest galaxy in our sample has
$r_e\approx 2$ pixels; the median effective radius is $r_e \approx 6$
pixels. Therefore, we anticipate that the net systematic effect of the
under-sampling on the fundamental plane is small, being 1-2\% on
average.

We caution against fitting profiles or growth curves too close to the
centers of galaxies, even when deconvolution of the WFPC2 images has
taken place, because the under-sampling makes fitting small galaxies
quite problematic. Techniques which improve the sampling, such as
``drizzling'' (\cite{driz}) should improve the usefulness of the data
for such objects.


\subsection{A Comparison of Stellar and Model Point-Spread Functions}

In both procedures, we have relied on the assumption that the Tiny Tim
point-spread function (\cite{krist}) is an accurate match to the PSF
of the observations. Even when this assumption is be correct for a
single exposure, the process of combining multiple exposures in each
pointing introduces extra ``jitter'' in the PSF of the resulting
image. The situation is actually worse because one has only a finite
number of images, {\it e.g.\/}, $N=3$, and not the limit of
$N\rightarrow \infty$ images. Therefore the individual jitter vectors
are not going to be isotropically distributed, for each galaxy (or
star) in a given pointing.

Therefore, we tested to what extent our results are effected by errors
in the model PSF. We built model galaxy images using a bright star as
the {\it observed\/} PSF, and then analyzed these images using a Tiny
Tim model PSF. The star was located at approximately $(550, 541)$ in
CCD $\#4$ of pointing $\#1$. We performed this test using observed
stellar and Tiny Tim PSFs with $1\Sec 5$ diameter.

We generated a $10\times $ subsampled model PSF at the integral pixel
location of the star, shifted and resampled it to the pixel size of
the WF CCDs, and then convolved with the charge-smearing kernel. We
measured the flux-weighted centroid of this nominally sampled PSF and
adjusted the sub-pixel shift until its centroid coincided with that of
the star. The effects due to under-sampling in the star and Tiny Tim
point-spread functions should cancel out, because they are positioned
at identical sub-pixel locations (to within 5\% of a pixel).

We created a set of 2D model galaxy images which had been convolved
with the stellar image. The models had pure $r^{1/4}$-law profiles
with effective radii $r_e \in (1,2,4,8,16)$ pixels. Using the
sub-pixel shifted Tiny Tim PSF, we then measured the structural
parameters of the model using the direct image fitting. The measured
effective radii underestimated the correct values by 17\%, 12\%, 8\%,
6\%, and 4\%, respectively. This suggests that the model PSF was
broader than the star, but because of the under-sampling, this
difference in width was not initially obvious. The fundamental plane
parameters were systematically larger by 6.8\%, 3.6\%, 1.8\%, 0.8\%,
and 0.2\%, respectively. These offsets are independent of model
ellipticity.

The smallest galaxy in our sample has an effective radius of $r_e
\approx 2$ pixels ($0\Sec 2$). The median half-light radius of the
sample is 0\Sec 6, with 25 and 75 percentiles of 0\Sec 35 and \about
$1''$. Thus, for the upper half of the sample, the uncertainties on
the fundamental plane parameters, due to improper matching of the
point-spread functions, are $\simlt 1\%$. For the 25\% of the sample
with $0\Sec 35 < r_e < 0\Sec 6$, the effects are $\simlt 3\%$. For the
remainder, the uncertainty is $< 4\%$. For galaxies at higher
redshift, or for galaxies farther down the luminosity function, one
must take extreme care to match the point-spread function to the data,
given the potentially large systematic uncertainties at $r_e\approx 1$
pixels.

\medskip

Using two distinct fitting methods, we have reliably derived effective
radii and surface brightnesses. The uncertainties in the effective
radii of the de Vaucouleurs profile fits are small, at the level of
\about 13\% \rms, with a mean offset of only 1\%. The measurements are
robust against fitting range, uncertainties in the sky background, and
fitting method. The uncertainties in the fundamental plane parameters
themselves is even smaller, \about 2.5\% \rms, with potential
systematic uncertainties at the level of $\simlt 3\%$, due to errors
in the PSF, under-sampling, and differences between the fitting
methods.

However, other sources of uncertainty in half-light radii may persist,
due to deviations in galaxy profiles from the $r^{1/4}$-law form.
Therefore, we continue with a discussion of systematic deviations from
the $r^{1/4}$-law.


\section{Sersic $r^{1/n}$-law Profiles}
\label{genprof}

The actual surface brightness profiles of galaxies often deviate from
pure $r^{1/4}$-laws. In the sample presented here, both bulge and disk
components are readily visible in many of the images (see Figures
\ref{images} and \ref{residuals}). We now investigate the fitting of
more generalized surface brightness profiles to our data to test (1)
if a more complex profile is better suited to the profiles of the
CL1358+62 galaxies; (2) if such departures from an $r^{1/4}$-law are
important for the determination of their structural parameters; and
(3) if they allow for accurate and objective morphological
classification.

In this section, we generalize the parameterization of the surface
brightness profiles to the $r^{1/n}$-law (\cite{caon,graham}), where
$n$, in the case of this paper, is restricted to the set of integers
$n \in \{1 \ldots 6\}$. Note that the case of $n=1$ corresponds to an
exponential disk, and $n=4$ is equal to the de Vaucouleurs profile.

The inclusion of $n$ serves two purposes. First, the best-fit $n$ is a
a quantitative descriptor of galaxy morphology. Graham \etal\ (1997)
found that the giant cluster ellipticals and BCGs follow $n>4$
profiles. Furthermore, Saglia \etal\ (1997) showed that $r^{1/n}$ is
equivalent to a subset of bulge-plus-disk models, though the
relationship between $n$ and $B/D$ is not a straightforward one.


\subsection{Fitting the Growth Curves}

Integrated $r^{1/n}$-laws, with $n=1,2,3,4,5,6$ have been fit to the
growth curves of \S \ref{growthfit}. The enclosed flux within a given
radius for an $r^{1/n}$-law profile can be simply expressed:
\begin{eqnarray}
F_n(r) &=& \pi r_n^2 \langle I\rangle_n {2\gamma(2n,x)\over (2n-1)!},\nl
\nonumber x &=& (2n-0.324) \biggl({r\over r_n}\biggr)^{1/n}
\label{analn}
\end{eqnarray}
where $r_n$ refers to the half-light radius of the corresponding
$r^{1/n}$-law fit. Fundamental plane parameters will be expressed as
$r_n I_n^{0.76}$ and $r_n \langle I\rangle_n^{0.76}$.

As in the case of the de Vaucouleurs fitting ($n=4$), we adopt
structural parameters derived from an inner fitting radius of 1 pixel,
and an outer radius of $2\,r_n$. We tested the structural parameters
from these fitting functions with direct image model fitting using the
$r^{1/n}$-laws for a restricted set of $n$ ($n=1 \ldots 4$).
Comparisons of the structural parameters derived from the image and
growth-curve fitting for $n=1,2,3$ were similar to the comparisons for
$n=4$ discussed in \S \ref{growthvsdirect} (in which differences in
$r_4 I_4^{0.76}$ were 1.5\% in the mean, with 2.5\% \rms\ scatter).


\subsection{A Comparison of the $r^{1/n}$ Profiles}

By expanding our growth curve fitting to a range of profile shapes,
one expects to gain additional knowledge about the galaxy sizes and
morphologies. In this section, we investigate whether or not the new
measures of half-light radii are necessarily more accurate than those
obtained with the de Vaucouleurs profile. Figure \ref{rrat}(a) shows
the distribution of the ratio $r_n/r_e$ (where $r_e = r_4$) using the
best-fit $n$ for a given galaxy. This distribution is quite broad,
suggesting that the fit of a pure de Vaucouleurs profile can lead to
large errors in half-light radius, similar to what has been found for
nearby samples (\cite{caon}).

\begin{figure}
\mkfigbox{4in}{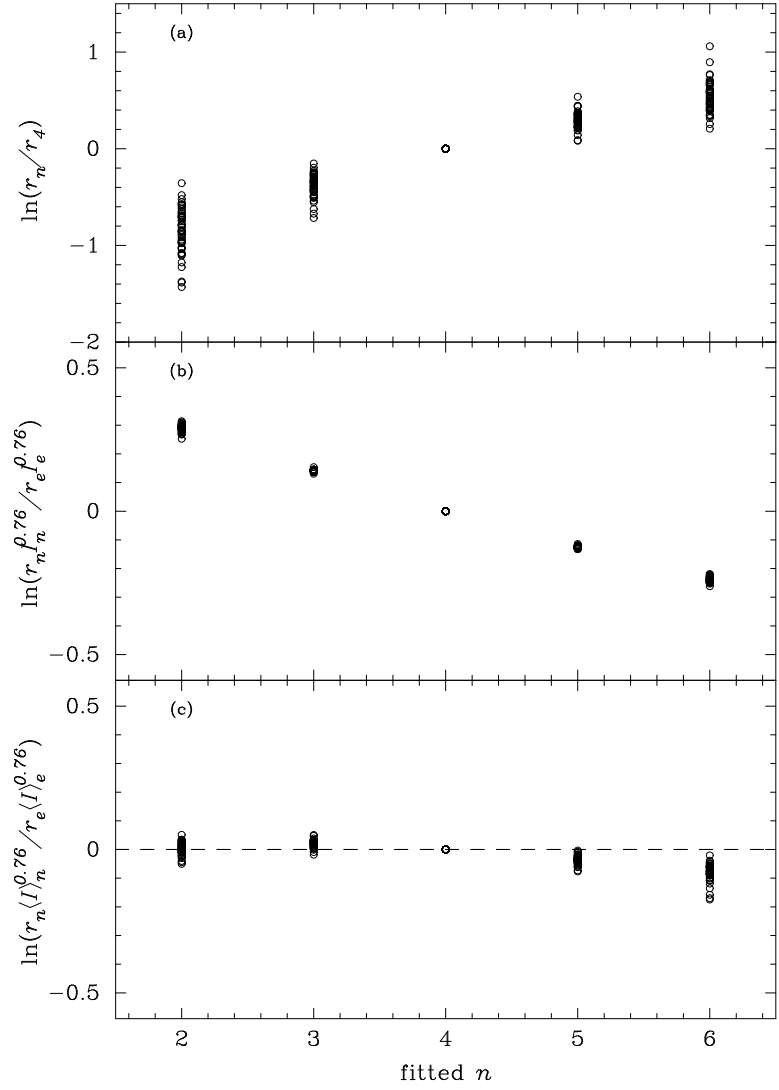}
\caption[Variation of $r_{1/2}$, and FP parameters as a function of $n$]{
(a) The ratio of half-light radii derived from all $n\ne 4$ to the
effective radius from the de Vaucouleurs ($n=4$) fit. (b) The ratio of
the effective fundamental plane parameter from these $n$ fits, with
the effective fundamental plane parameter from the de Vaucouleurs fit.
(c) The ratio of the mean effective fundamental plane parameter from
these $n$ fits, with the mean effective fundamental plane parameter
from the de Vaucouleurs fit. The data for all 53 galaxies are shown,
for every value of $n$ fit.
\label{sfn}}
\end{figure}

For a given $n$, the ratio of $r_n/r_4$ appears to be a well defined
function. As shown in Figure \ref{sfn}(a), $\ln(r_n/r_4) \approx
-1.32+0.33n$, almost irrespective of the true shape of the galaxies,
with a scatter of less \about 5\%. Thus, the fit of an $n=6$ profile
to an intrinsically $n=2$ galaxy will find too large a half-light
radius by nearly a factor of three. As a result, a large bulk of an
$n=6$ profile is fit to parts of the growth curve which are more
susceptible to errors in the sky subtraction. In crowded regions,
where a large source of systematic error is the background
determination, residual background light can easily appear as a
shallow envelope in the integrated surface brightness profiles,
masquerading as the profile of an $r^{1/6}$-law, or worse yet, even
larger values of n. Of course, the other extreme is also possible. By
over-estimating the sky, one can effectively decrease the $n$ derived
for a given galaxy. Therefore, when expanding the profile fitting to
include the overall shape ($n$), one must treat the background
subtraction with care.

While for fixed values of $n$, the individual parameters of $r_n$ and
$I_n$ are not particularly sensitive to errors in background
subtraction (see the earlier case of $n=4$), the chief uncertainty is
{\it which\/} value of $n$ to assign to a given galaxy. By varying the
range of radii over which one compares the different $r^{1/n}$ fits,
one begins to test the sensitivity of $n$ to the uncertainties in sky
subtraction. A 3\% error in sky subtraction (\S \ref{deriv}) can lead
to changes in $n$, $\Delta n \approx 1$, and errors in half-light
radius of 20-30\%. The typical scatter introduced in $n$ due to the
sky extrapolation alone is of order $\Delta n\approx \pm 0.2$. Nearly
a third of the sample, however, have $\Delta n$ values larger than
this, and many of these are the $n\simgt 4$ galaxies for which
uncertainties in $n$ on the order of unity are not uncommon. We adopt
$\Delta n$ as the uncertainty in the measurement of $n$, and the
faintest galaxies, as expected, tend to have the largest
uncertainties. We conclude that the $r^{1/n}$-law does not yield more
accurate estimates of half-light radii, especially for galaxies fit by
$n> 4$.

Several authors have commented upon a trend of galaxy shape parameter,
``$n$,'' with galaxy size (\cite{caon,graham}). However, in Figure
\ref{sfn}(a) we saw that the ratio of $r_n/r_4$ scales in a
well-defined manner. By fitting a low-$n$ profile to a galaxy, one
will automatically find a smaller half-light radius than when fitting
a larger $n$ profile. The half-light radii from low-$n$ fits are
simply smaller than the half-light radii one derives when fitting
larger-$n$ profiles, independent of the true shape of the galaxy
profile. This bias may complicate the nature of the trend of $n$ with
galaxy size.


\subsection{Effective and Mean Effective Surface Brightness}
\label{surf}

Ideally, the fundamental plane should be insensitive to the detailed
profiles of (early-type) galaxies. However, the fundamental plane
parameter constructed with the surface brightness at an effective
radius is not robust. As can be seen from Figure \ref{sfn}(b), the
ratio of $r_1 I_1^{0.76}$ to $r_6 I_6^{0.76}$ is a factor of
\about$2.5$. The ratios of the $r_n I_n^{0.76}/r_m I_m^{0.76}$ to each
other are {\it independent of the real shapes of the galaxy
profiles\/}, to better than a few percent. In Figure \ref{sfn}(c),
however, we see that the fundamental plane parameter constructed with
the {\it mean surface brightness\/} (within $r_n$) is much more robust
with respect to $n$, showing only $\about 10\%$ peak-to-peak
difference in $r_{n}\langle I\rangle_{n}^{0.76}$. The typical scatter
about this relation is only a few percent.

The ratio of $r_n \langle I\rangle_n^{0.76}$ to $r_4 \langle I
\rangle_4^{0.76}$ is virtually independent of galaxy profile as well.
This indicates that every $n=2$ galaxy fit by a de Vaucouleurs profile
is actually measured incorrectly in precisely the same way as every
other $n=2$ galaxy, independent of galaxy size.

\medskip

In summary, the fitting of the Sersic $r^{1/n}$-law profiles does not
lead to more accurate photometric parameters for the purposes of
fundamental plane analysis. The derived half-light radii are
correlated with the best-fit $n$, which is sensitive to errors in the
sky background. Fortunately, the fundamental plane is insensitive to
these uncertainties in the photometric parameters, because the
fundamental plane parameters, when constructed using mean effective
surface brightnesses, remain constant at the level of 3\%.


\section{Bulge/Disk Decompositions and Quantitative Measures of
Morphology}
\label{scomp}

Classifying galaxies by morphology becomes increasingly difficult at
high redshifts. Distant galaxies are typed using images taken with
(linear) CCD detectors, while nearby samples have been primarily typed
using (non-linear) photographic plates. Furthermore, the distant
galaxies typically span fewer resolution elements, and the images have
poorer $S/N$ than images of the nearby prototype elliptical and
lenticular galaxies with which we wish to make comparisons. In order
to maximize the use of our data in determining galaxy morphologies, we
classify the galaxies in our sample by both visual and quantitative
morphology. The visual classifications have been taken from Fabricant
\etal\ (1999).

In the previous section, we derived the $n$ parameters for the
galaxies in our sample. As stated earlier, the Sersic (1968) $n$
profiles are a ``restricted'' subset of bulge-plus-disk models
(\cite{saglia}). In fitting the growth curves, the use of $n$ as a
descriptor of morphology is formally more robust than bulge-disk
decompositions. One needs only three quantities to describe a profile,
$(n,r_e,I_e)$, rather than four, $(r^B_{1/2}, I^B_{1/2}, r^D_{1/2},
I^D_{1/2})$.

In this section, we test whether $n$ reasonably reflects $B/D$ ratios
for our sample of galaxies. To this end, two-component,
bulge-plus-disk growth curves were fit to the integrated surface
brightness profiles of our galaxies, using the same fitting radii as
above. The growth curves were fit between 1 pixel ($0\Sec 1$) and $2\,
r_n$, defined by the ``best'' fit $r^{1/n}$-law (above). As in the
fitting of the $r^{1/4}$-law growth curves, we used the
Levenberg-Marquardt method to search for the parameters which
minimized $\chi^2$, taking extra care to avoid local minima.

In Figure \ref{profile}, we show the fitted bulge-plus-disk
superpositions and the residuals from this fit using thin solid lines.
As is shown in the figures, the galaxies in our sample appear to be
well described by the superpositions of de Vaucouleurs bulge ($n=4$)
and exponential disk ($n=1$) profiles. The individual bulge and disk
components are plotted as dashed and dotted lines, respectively. Note
that the bulge-plus-disk models tend to provide better fits than a
pure $r^{1/4}$-law. The addition of an exponential disk improves the
fit for many of the galaxies. The E+A galaxies, ID \# 209, 328, and
343, are bulge-dominated, but clearly have disks at the level of
10-35\%. Over the entire sample, the median bulge fraction is 84\%.
Upon visual inspection of Figure \ref{residuals}, one finds disk-like
structures in about half of the sample, consistent with the visual
classifications. Spiral arms are also plainly visible in many,
including two of the E+A galaxies.

In many of the galaxies, there appears to be evidence for central
exponential (disk) structures. Most of the galaxies with larger bulge
fractions show evidence for central exponential components in their
surface brightness profiles. This does not appear to be an artifact of
the fitting procedures or errors in the PSF, as many of these central
``disk'' components persist at radii $r\gg 1$ pixel. Evidence of such
structures might appear in the kinematic profiles but seeing would
complicate the interpretation of such data (\cite{kelson99a}). A full
study of the entire membership catalog using two-dimensional model
fitting of the images (\cite{tran}), would help confirm whether this
abundance of central exponentials is the norm, or an artifact of the
deconvolution process.

Table \ref{params} lists the optical morphologies (\cite{dgf99}) with
the best-fit $n$ parameters. We list (1) galaxy ID, (2) visual
morphology, $T$, from Fabricant \etal\ (1999), (3) $(B-V)$ color, also
from van Dokkum \etal\ (1998a), (4) $r_e$, (5) $\langle \mu\rangle
_e$, (6) $r_e\langle I\rangle_e^{0.76}$, (7) $n$, (8) $r_{1/2}^{n}$,
(9) $\langle \mu\rangle _{1/2}^{n}$, and (10) $r_n\langle
I\rangle_n^{0.76}$. The fundamental plane parameters have been
expressed using units of kpc and $L_{\odot}/$pc$^2$ ($H_0=65$
\kms\Mpc-1, $q_0=0.1$).

Table \ref{bdstruct} gives the best-fit bulge fractions for the
galaxies in this sample, and lists the individual values of (3)
$r_{1/2}$, (4) $\langle \mu\rangle_{1/2}$, (5) $r_{1/2}\langle
I\rangle_{1/2}^{0.76}$, (6) $r_{1/2}$ for the bulge component, (7)
$\langle \mu\rangle _{1/2}$ for the bulge component, (8) $r_{1/2}$ for
the disk component, (9) $\langle \mu\rangle _{1/2}$ for the disk
component, Bulge Fraction within (10) $r\le 0\Sec 2$, (11) $r\le 0\Sec
5$, and (12) $r\le 2\,r_{1/2}$, (13) total bulge fraction, and (14)
luminosity-weighted ellipticity, are given for each galaxy. Scale
lengths are listed in arcsec, and surface brightnesses in $V_z\ \rm
mag/\square ''$. The tables include only formal errors and do not
include other sources of errors, such as uncertainties in the
calibration or errors due to PSF mismatch.

\begin{figure}
\mkfigbox{8.8cm}{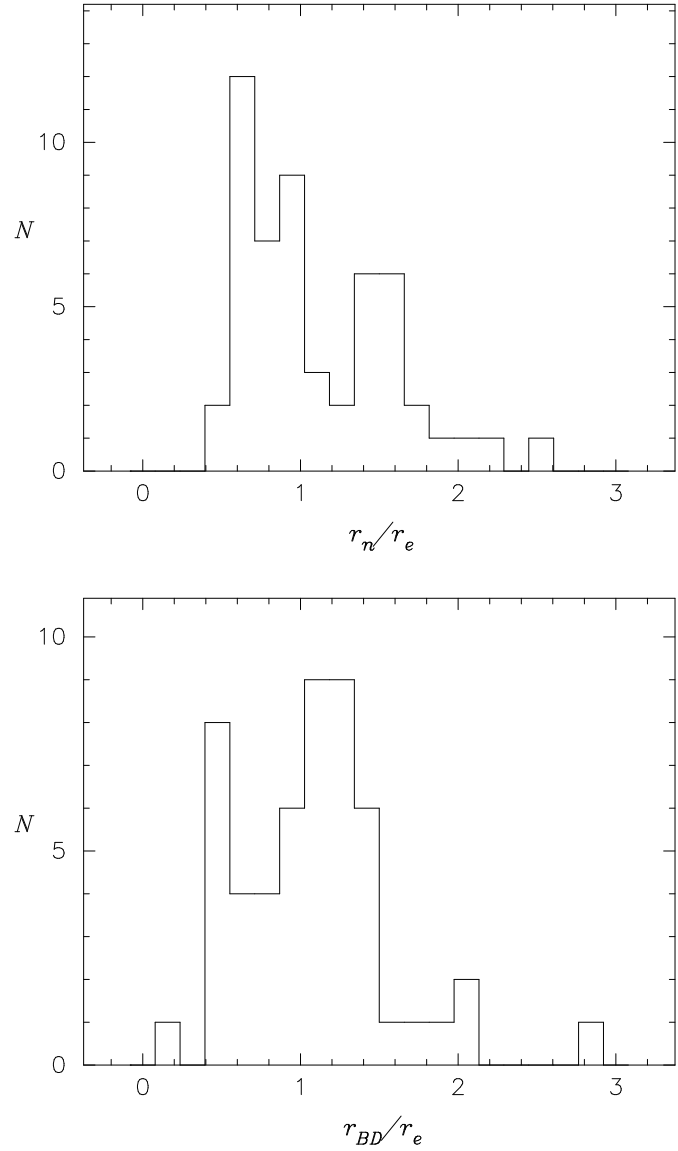}
\caption[Differences in half-light radii from de Vaucouleurs, $r^{1/n}$,
and $B/D$ profiles]{
(a) The distribution of the ratio of half-light radii, derived from
the best-fit $r^{1/n}$-laws, to the half-light radii from the de
Vaucouleurs ($n=4$) fit. (b) The distribution of the ratio of
half-light radii, derived from the bulge-plus-disk fits, to the
half-light radii from the de Vaucouleurs fit.
\label{rrat}}
\end{figure}

There were two explicit reasons for expanding the form of the fitted
surface brightness profiles to the $r^{1/n}$-laws and bulge-plus-disk
superpositions; they are to (1) determine accurate measures of
half-light radii; and (2) objective morphological classifications.
Earlier we compared the half-light radii from the Sersic profiles to
those derived using the de Vaucouleurs profile. In Figure
\ref{rrat}(b), we show the distribution of bulge-plus-disk half-light
radius, $r_{B+D}$, divided by $r_e$, the half-light radius of the de
Vaucouleurs profile fit. The distribution is broad, and with Figure
\ref{rrat}(a), should further caution the reader about the reliability
of half-light radii in general.

\begin{figure}
\mkfigbox{5in}{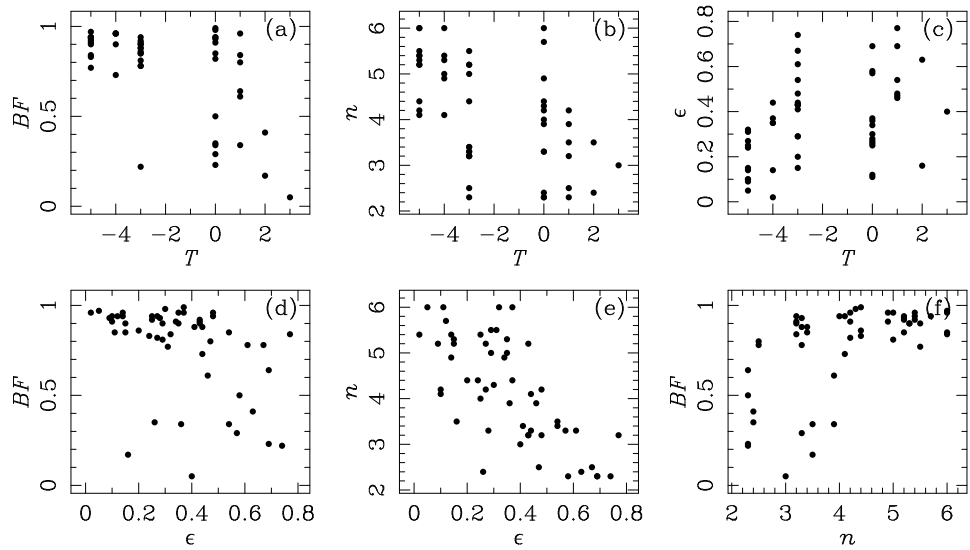}
\caption[Ratio of bulge radii to disk radii]{
(a-c) Bulge fraction, $n$, and ellipticity are plotted against the
visual classifications from Fabricant \etal\ (1999). Note that, as
expected, visually classified early-types have high bulge fractions
and typically have large values of $n$. In (d) and (e) we plot the
bulge fraction and $n$ against ellipticity. In (f) we compare $n$ with
$BF$ and note that those galaxies which are best-fit using large
values of $n$ tend to have high bulge fractions. 
\label{compare}}
\end{figure}

The other reason behind expanding the model surface brightness
profiles to Sersic profiles, and bulge-plus-disk superpositions was to
test whether these forms could be used to objectively classify the
galaxies, given the recent interest in quantitative morphologies. In
Figure \ref{compare}, we compare visual and quantitative estimates of
morphology to help test their reliability. Figures \ref{compare}(a-c)
compare $BF$, the bulge fraction of total light, $n$, the best-fit
$r^{1/n}$-law, and $\epsilon$, the flux-weighted apparent
ellipticities from the image modeling to the visual classifications,
$T$, of Fabricant \etal\ (1999). In general, the visually classified
E/S0s ($T\le -3$) have high bulge fractions. Those galaxies with $n\ge
4$ are all of type S0/a and earlier. While $T$ and $n$ are clearly
correlated, the relationship between profile shape, $n$, and visual
morphology has large scatter. Figure \ref{compare}(d,e) shows $BF$ and
$n$ as functions of $\epsilon$. In Figure \ref{compare}(f) one sees
that galaxies with $n\simgt 4$ typically have high bulge fractions.
Clearly the relationship between bulge fraction, $n$, and optical
morphology is not straightforward. That there appears to be no tight
relation between $n$ and $BF$, however, should not be surprising.
Saglia \etal\ (1997) showed that $n$ is a complicated combination of
the bulge fraction and the ratio of disk to bulge half-light radii. We
conclude that there is only a level of correspondence between the
different estimates of morphology which allows one to discriminate
between bulge-dominated and disk-dominated galaxies. It is clear from
the figures that these quantitative measures are not successful at
distinguishing subtle differences between adjacent morphological
classes ({\it e.g.\/}, Es from S0s).


\section{Summary of the Structural Parameters}
\label{adopt}

The primary goal of this paper is to provide the photometric
parameters required to compare the fundamental plane in CL1358+62 to
that measured locally. We have shown that the growth-curve structural
parameters are consistent with those derived from directly fitting 2D
models to the images.

However, the direct image fitting implemented by van Dokkum \& Franx
(1996) is currently restricted to rather simple galaxy models: a
single component with constant ellipticity and position angle. Other,
more sophisticated algorithms, such as those described by Simard
\etal\ (1997), could be used as well. Given the magnitudes of the
uncertainties in half-light radii (cf. Figure \ref{rrat}), the more
sophisticated methods will not likely result in more accurate
half-light radii, but simply in greater accuracy in the estimation of
their errors. More distant galaxies, with smaller angular sizes, will
require more sophisticated modeling to avoid serious systematic errors
in structural parameters due to the under-sampling. Despite these
issues, we have shown that the measurements of $r_e \langle
I\rangle_e^{0.76}$ are quite robust.

The current errors in the structural parameters, as deduced by the
differences in the results of our two methods, give us confidence that
more complicated procedures would not greatly improve on the analysis.
What we do find, however, is that one can use $n$ to {\it grossly\/}
distinguish between bulge- and disk-dominated systems. These
$r^{1/n}$-laws, however, do not give much insight into any particular
physical properties of the galaxies, such as ratios of bulge to disk
radii, or bulge fraction.

Given the uncertainties in the actual profile shapes of the CL1358+62
galaxies, the fundamental plane parameter must be constructed with
care. Use of the surface brightness at the half-light radius appears
to be a strong function of the profile chosen for the fitting (see
Figure \ref{sfn}). Adopting the mean surface brightness within the
half-light radius produces a fundamental plane parameter which is
robust against the adopted profile shape.

\begin{figure}
\mkfigbox{3in}{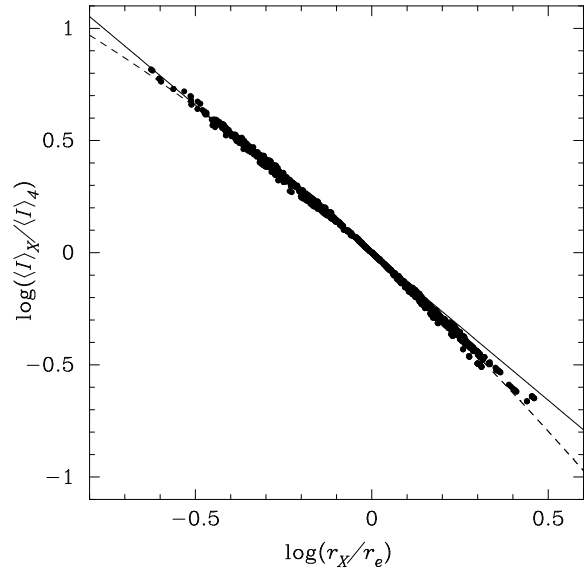}
\caption[Error correlation]{
For every galaxy in the sample, we show the ratios of $\langle
I\rangle_X/ \langle I\rangle_4$ vs. $r_X/r_4$, where $X$ denotes
either the $r^{1/n}$-law structural parameters ($n=1\ldots 6$) or the
bulge-plus-disk structural parameters. Note that $r_e \equiv r_4$. For
this plot, the structural parameters derived from every tested fitting
range are used, thus several points exist for each galaxy and profile
shape. The solid line indicates the curve of constant $r\langle
I\rangle^{0.76}$, running parallel to the fundamental plane. The
dashed line is the curve expected for a pure $r^{1/4}$-law growth
curve.
\label{errcor}}
\end{figure}

Although we have outlined potentially large uncertainties in
half-light radii, the correlation between $r$ and $\langle I\rangle$
ensures that the fundamental plane parameter, $r\langle
I\rangle^{0.76}$ is conserved, with a scatter of 3\%. In Figure
\ref{errcor}, we show the correlation between $\langle I\rangle$ and
$r$ explicitly for the entire sample. For every trial fitting range
and $r^{1/n}$-law, we plot the mean surface brightness within the
fitted half-light radius, normalized by the adopted mean effective
surface brightness from the de Vaucouleurs profile fit, versus galaxy
radius, normalized by the effective radius from the de Vaucouleurs
profile fit. The data do not deviate significantly from the
correlation expected from pure $r^{1/4}$-law growth curves, shown by
the dashed line. The slope running parallel to the fundamental plane,
$r\langle I\rangle^{0.76} = {\rm constant}$, is shown by the solid
line. Given the large uncertainties in half-light radii, the strong
error correlation may not only be coincidentally parallel to the
fundamental plane, but also likely seriously biases the measured slope
of the scaling relation (\cite{thesis,models}).

Because the analysis most closely parallels earlier, ground-based
work, we will use the results from the growth curve fitting in the
derivation of the fundamental plane (see \cite{kelson99b}).


\section{Conclusions}
\label{conclusions}

Using deep HST imaging, we have extracted growth curves for a sample
of 53 galaxies in the cluster CL1358+62 at $z=0.33$. These growth
curves have been fit using integrated $r^{1/4}$-laws, integrated
Sersic $r^{1/n}$-law profiles, and bulge-plus-disk superpositions.
Comparisons of the fits between the different parameterizations of the
profiles show an \rms\ scatter of 30-40\% in half-light radii,
although the formal uncertainties in the half-light radii from a given
fit are typically smaller by factors of three. The half-light radii
derived from the Sersic profiles agree in the median with those found
using the $r^{1/4}$-laws at the level of 1\%. The half-light radii
derived using the bulge-plus-disk superpositions are systematically
larger by 10\%. Despite the potentially large uncertainties in the
half-light radii, the quantity that enters the fundamental plane,
$r_n\langle I\rangle_n^{0.76}$ has been shown to be quite stable. This
combination is shown to be nearly parallel to the correlation between
radius and mean surface brightness for an $r^{1/4}$-law growth curve.

The median half-light radius of this sample is 0\Sec 61 (2.9 kpc,
$H_0=65$\kms\Mpc-1, $q_0=0.1$), with a range of 1.26 dex in $r_e$.
(Using $B/D$ decompositions, or $r^{1/n}$-laws, only changes the
median by $\pm0\Sec 01$.) The range is smaller than the 1.6 dex
spanned by the J\o{}rgensen \etal\ (1993) Coma sample, probably due to
differences in selection criteria. 

The median bulge fraction is about 84\%. Many of the galaxies with
larger bulge fractions show evidence for central exponential
components in their surface brightness profiles. The few E+A galaxies
in this sample invariably have disks (see also \cite{franx98}. Their
bulge fractions are \about 65-90\% (see also, {\it e.g.\/},
\cite{wirth}). Half of the sample shows evidence for disky structures
after subtracting two-dimensional $r^{1/4}$-law models from the galaxy
images. This result is consistent with the visual classifications in
which nearly half of the sample are of type S0/a trough Sb. Two of the
three E+A galaxies have evidence of spiral arms in their residual
maps.

In the future, we plan to derive accurate structural parameters for
the entire sample of confirmed cluster members and investigate the
global correlations of these parameters with the signatures of stellar
populations (color), and environment (clustercentric radius and local
densities) (\cite{tran}). The importance of these issues for the
fundamental plane is discussed by Kelson \etal\ (1999b). Studies of
galactic structural parameters and scaling relations in distant
clusters, like this program, clearly require wide-field imaging of the
distant cluster fields. Two more WFPC2 mosaics have now been taken, of
MS2053-04 ($z=0.58$) and MS1054--03 ($z=0.83$). With these clusters,
the evolution of the properties of cluster galaxies can be accurately
mapped to redshifts approaching unity and the Advanced Camera for
Surveys is expected to extend such studies to redshifts beyond $z=1$.

\acknowledgements

We gratefully acknowledge D. Koo and S. Faber, who provided valuable
comments on an early version of the paper. Furthermore, we appreciate
the effort of all those in the HST program that made this unique
Observatory work as well as it does. The assistance of those at STScI
who helped with the acquisition of the HST data is also gratefully
acknowledged. Support from STScI grants GO05989.01-94A,
GO05991.01-94A, and AR05798.01-94A and NSF grant AST-9529098 is
gratefully acknowledged.


\newpage

\begin{deluxetable}{r l c r r c r r r r c}
\tablewidth{0pt}
\tiny
\tablecaption{1D $r^{1/4}$- and $r^{1/n}$-law Structural
Parameters\label{params}}
\tablehead{
\colhead{}&
\colhead{}&
\colhead{}&
\multicolumn{3}{c}{$r^{1/4}$-law}&
\colhead{}&
\multicolumn{3}{c}{$r^{1/n}$-law}\nl
\colhead{ID}&
\colhead{Type}&
\colhead{$(B-V)_z$}&
\colhead{$r_e$}&
\colhead{$\langle \mu\rangle_e$}&
\colhead{$r_e \langle I\rangle_e^{0.76}$}&
\colhead{}&
\colhead{$n$}&
\colhead{$r_{n}$}&
\colhead{$\langle \mu\rangle_{n}$}&
\colhead{$r_n \langle I\rangle_n^{0.76}$}\nl
\colhead{(1)}&
\colhead{(2)}&
\colhead{(3)}&
\colhead{(4)}&
\colhead{(5)}&
\colhead{(6)}&
\colhead{}&
\colhead{(7)}&
\colhead{(8)}&
\colhead{(9)}&
\colhead{(10)}\nl
}
\startdata
095&S0  & 0.894& 0.356&  19.84&   2.59&& 5.2& 0.565&  20.59&   2.56\nl
$\ldots$&$\ldots$&$\ldots$&$\pm 0.004$&$\pm   0.02$&$\ldots$&&$\pm 1.2$&$\pm 0.005$&$\pm   0.02$&$\ldots$\nl
110&S0/a& 0.850& 0.648&  20.32&   2.70&& 2.3& 0.375&  19.53&   2.71\nl
$\ldots$&$\ldots$&$\ldots$&$\pm 0.012$&$\pm   0.04$&$\ldots$&&$\pm 0.4$&$\pm 0.005$&$\pm   0.03$&$\ldots$\nl
129&S0/a& 0.830& 0.324&  19.27&   2.72&& 4.2& 0.326&  19.28&   2.72\nl
$\ldots$&$\ldots$&$\ldots$&$\pm 0.004$&$\pm   0.02$&$\ldots$&&$\pm 0.3$&$\pm 0.004$&$\pm   0.03$&$\ldots$\nl
135&S0  & 0.827& 0.351&  20.19&   2.48&& 5.2& 0.493&  20.71&   2.47\nl
$\ldots$&$\ldots$&$\ldots$&$\pm 0.009$&$\pm   0.05$&$\ldots$&&$\pm 0.3$&$\pm 0.010$&$\pm   0.04$&$\ldots$\nl
142&S0/a& 0.856& 1.414&  22.15&   2.49&& 3.9& 1.398&  22.13&   2.49\nl
$\ldots$&$\ldots$&$\ldots$&$\pm 0.046$&$\pm   0.07$&$\ldots$&&$\pm 0.9$&$\pm 0.046$&$\pm   0.07$&$\ldots$\nl
164&S0/a& 0.870& 1.165&  21.56&   2.58&& 5.7& 1.955&  22.37&   2.56\nl
$\ldots$&$\ldots$&$\ldots$&$\pm 0.035$&$\pm   0.06$&$\ldots$&&$\pm 0.7$&$\pm 0.100$&$\pm   0.08$&$\ldots$\nl
182&S0  & 0.839& 0.418&  20.70&   2.40&& 3.2& 0.310&  20.25&   2.40\nl
$\ldots$&$\ldots$&$\ldots$&$\pm 0.008$&$\pm   0.04$&$\ldots$&&$\pm 0.3$&$\pm 0.005$&$\pm   0.04$&$\ldots$\nl
209&Sa  & 0.601& 0.980&  21.34&   2.57&& 2.3& 0.557&  20.49&   2.59\nl
$\ldots$&$\ldots$&$\ldots$&$\pm 0.018$&$\pm   0.04$&$\ldots$&&$\pm 0.3$&$\pm 0.005$&$\pm   0.02$&$\ldots$\nl
211&S0  & 0.870& 0.612&  20.58&   2.60&& 3.3& 0.470&  20.18&   2.61\nl
$\ldots$&$\ldots$&$\ldots$&$\pm 0.015$&$\pm   0.05$&$\ldots$&&$\pm 0.4$&$\pm 0.009$&$\pm   0.04$&$\ldots$\nl
212&E   & 0.868& 0.547&  20.44&   2.59&& 5.2& 0.781&  21.01&   2.58\nl
$\ldots$&$\ldots$&$\ldots$&$\pm 0.009$&$\pm   0.03$&$\ldots$&&$\pm 0.3$&$\pm 0.008$&$\pm   0.02$&$\ldots$\nl
215&S0  & 0.843& 0.338&  20.17&   2.47&& 5.0& 0.460&  20.67&   2.45\nl
$\ldots$&$\ldots$&$\ldots$&$\pm 0.006$&$\pm   0.03$&$\ldots$&&$\pm 1.0$&$\pm 0.007$&$\pm   0.03$&$\ldots$\nl
233&E/S0& 0.847& 0.590&  19.95&   2.78&& 5.3& 0.873&  20.58&   2.75\nl
$\ldots$&$\ldots$&$\ldots$&$\pm 0.012$&$\pm   0.04$&$\ldots$&&$\pm 0.4$&$\pm 0.010$&$\pm   0.03$&$\ldots$\nl
234&Sb  & 0.689& 3.636&  23.68&   2.43&& 3.0& 2.264&  22.97&   2.44\nl
$\ldots$&$\ldots$&$\ldots$&$\pm 0.349$&$\pm   0.18$&$\ldots$&&$\pm 1.0$&$\pm 0.166$&$\pm   0.16$&$\ldots$\nl
236&S0  & 0.895& 0.539&  20.51&   2.57&& 5.5& 0.892&  21.33&   2.53\nl
$\ldots$&$\ldots$&$\ldots$&$\pm 0.009$&$\pm   0.03$&$\ldots$&&$\pm 1.5$&$\pm 0.010$&$\pm   0.03$&$\ldots$\nl
242&E   & 0.882& 1.201&  21.54&   2.60&& 4.2& 1.210&  21.55&   2.60\nl
$\ldots$&$\ldots$&$\ldots$&$\pm 0.026$&$\pm   0.04$&$\ldots$&&$\pm 0.3$&$\pm 0.026$&$\pm   0.04$&$\ldots$\nl
256&E   & 0.875& 0.956&  20.30&   2.88&& 5.4& 1.542&  21.06&   2.86\nl
$\ldots$&$\ldots$&$\ldots$&$\pm 0.010$&$\pm   0.02$&$\ldots$&&$\pm 0.4$&$\pm 0.010$&$\pm   0.02$&$\ldots$\nl
269&E/S0& 0.913& 0.776&  20.05&   2.86&& 5.0& 1.056&  20.55&   2.85\nl
$\ldots$&$\ldots$&$\ldots$&$\pm 0.008$&$\pm   0.02$&$\ldots$&&$\pm 1.0$&$\pm 0.008$&$\pm   0.01$&$\ldots$\nl
292&S0/a& 0.776& 0.911&  21.21&   2.58&& 2.4& 0.562&  20.50&   2.59\nl
$\ldots$&$\ldots$&$\ldots$&$\pm 0.031$&$\pm   0.07$&$\ldots$&&$\pm 0.5$&$\pm 0.009$&$\pm   0.04$&$\ldots$\nl
298&S0  & 0.914& 0.642&  19.93&   2.82&& 2.5& 0.406&  19.23&   2.83\nl
$\ldots$&$\ldots$&$\ldots$&$\pm 0.007$&$\pm   0.02$&$\ldots$&&$\pm 0.5$&$\pm 0.002$&$\pm   0.01$&$\ldots$\nl
300&S0  & 0.879& 0.386&  19.59&   2.70&& 3.4& 0.310&  19.25&   2.71\nl
$\ldots$&$\ldots$&$\ldots$&$\pm 0.004$&$\pm   0.02$&$\ldots$&&$\pm 0.4$&$\pm 0.002$&$\pm   0.02$&$\ldots$\nl
303&E   & 0.858& 0.494&  20.18&   2.63&& 5.3& 0.729&  20.81&   2.61\nl
$\ldots$&$\ldots$&$\ldots$&$\pm 0.007$&$\pm   0.03$&$\ldots$&&$\pm 0.4$&$\pm 0.007$&$\pm   0.02$&$\ldots$\nl
309&E/S0& 0.900& 0.433&  19.80&   2.69&& 4.9& 0.572&  20.23&   2.68\nl
$\ldots$&$\ldots$&$\ldots$&$\pm 0.003$&$\pm   0.01$&$\ldots$&&$\pm 1.0$&$\pm 0.003$&$\pm   0.01$&$\ldots$\nl
328&Sa  & 0.629& 0.878&  21.15&   2.58&& 2.5& 0.550&  20.45&   2.59\nl
$\ldots$&$\ldots$&$\ldots$&$\pm 0.016$&$\pm   0.04$&$\ldots$&&$\pm 0.6$&$\pm 0.008$&$\pm   0.02$&$\ldots$\nl
335&S0/a& 0.859& 0.512&  20.79&   2.46&& 4.4& 0.551&  20.89&   2.46\nl
$\ldots$&$\ldots$&$\ldots$&$\pm 0.009$&$\pm   0.04$&$\ldots$&&$\pm 0.4$&$\pm 0.009$&$\pm   0.04$&$\ldots$\nl
343&S0  & 0.670& 0.438&  20.66&   2.43&& 3.3& 0.353&  20.33&   2.44\nl
$\ldots$&$\ldots$&$\ldots$&$\pm 0.010$&$\pm   0.05$&$\ldots$&&$\pm 0.4$&$\pm 0.007$&$\pm   0.05$&$\ldots$\nl
353&E/S0& 0.889& 0.958&  20.79&   2.73&& 5.4& 1.495&  21.52&   2.70\nl
$\ldots$&$\ldots$&$\ldots$&$\pm 0.007$&$\pm   0.01$&$\ldots$&&$\pm 0.5$&$\pm 0.008$&$\pm   0.01$&$\ldots$\nl
356&Sa  & 0.889& 0.815&  20.62&   2.71&& 3.9& 0.741&  20.48&   2.71\nl
$\ldots$&$\ldots$&$\ldots$&$\pm 0.011$&$\pm   0.03$&$\ldots$&&$\pm 0.2$&$\pm 0.011$&$\pm   0.02$&$\ldots$\nl
359&S0  & 0.851& 0.475&  19.95&   2.68&& 4.4& 0.527&  20.12&   2.68\nl
$\ldots$&$\ldots$&$\ldots$&$\pm 0.005$&$\pm   0.02$&$\ldots$&&$\pm 0.4$&$\pm 0.005$&$\pm   0.02$&$\ldots$\nl
360&E   & 0.861& 0.241&  19.65&   2.48&& 5.5& 0.373&  20.37&   2.45\nl
$\ldots$&$\ldots$&$\ldots$&$\pm 0.004$&$\pm   0.03$&$\ldots$&&$\pm 0.6$&$\pm 0.009$&$\pm   0.04$&$\ldots$\nl
366&S0/a& 0.863& 0.587&  20.68&   2.55&& 3.3& 0.448&  20.28&   2.56\nl
$\ldots$&$\ldots$&$\ldots$&$\pm 0.008$&$\pm   0.03$&$\ldots$&&$\pm 0.4$&$\pm 0.005$&$\pm   0.03$&$\ldots$\nl
368&Sab & 0.862& 1.055&  22.09&   2.38&& 2.4& 0.649&  21.34&   2.39\nl
$\ldots$&$\ldots$&$\ldots$&$\pm 0.032$&$\pm   0.06$&$\ldots$&&$\pm 2.6$&$\pm 0.010$&$\pm   0.04$&$\ldots$\nl
369&S0/a& 0.879& 0.935&  21.70&   2.44&& 4.3& 0.985&  21.78&   2.44\nl
$\ldots$&$\ldots$&$\ldots$&$\pm 0.023$&$\pm   0.05$&$\ldots$&&$\pm 0.4$&$\pm 0.023$&$\pm   0.05$&$\ldots$\nl
371&Sa  & 0.871& 1.699&  21.84&   2.66&& 3.5& 1.456&  21.61&   2.66\nl
$\ldots$&$\ldots$&$\ldots$&$\pm 0.033$&$\pm   0.04$&$\ldots$&&$\pm 1.5$&$\pm 0.018$&$\pm   0.03$&$\ldots$\nl
372&Sa  & 0.867& 1.595&  22.09&   2.56&& 3.2& 1.191&  21.64&   2.57\nl
$\ldots$&$\ldots$&$\ldots$&$\pm 0.064$&$\pm   0.08$&$\ldots$&&$\pm 1.2$&$\pm 0.034$&$\pm   0.06$&$\ldots$\nl
375&E   & 0.894& 3.813&  22.44&   2.83&& 6.0& 7.498&  23.53&   2.79\nl
$\ldots$&$\ldots$&$\ldots$&$\pm 0.075$&$\pm   0.04$&$\ldots$&&$\pm 0.2$&$\pm 0.231$&$\pm   0.06$&$\ldots$\nl
381&E/S0& 0.866& 0.392&  19.92&   2.61&& 6.0& 0.756&  20.97&   2.57\nl
$\ldots$&$\ldots$&$\ldots$&$\pm 0.008$&$\pm   0.04$&$\ldots$&&$\pm 0.2$&$\pm 0.020$&$\pm   0.05$&$\ldots$\nl
397&S0/a& 0.851& 0.727&  21.29&   2.46&& 3.3& 0.472&  20.71&   2.45\nl
$\ldots$&$\ldots$&$\ldots$&$\pm 0.020$&$\pm   0.06$&$\ldots$&&$\pm 0.4$&$\pm 0.011$&$\pm   0.04$&$\ldots$\nl
408&S0  & 0.861& 0.353&  19.70&   2.63&& 3.4& 0.291&  19.41&   2.63\nl
$\ldots$&$\ldots$&$\ldots$&$\pm 0.003$&$\pm   0.02$&$\ldots$&&$\pm 0.5$&$\pm 0.002$&$\pm   0.02$&$\ldots$\nl
409&E   & 0.895& 0.536&  21.38&   2.30&& 4.1& 0.525&  21.35&   2.30\nl
$\ldots$&$\ldots$&$\ldots$&$\pm 0.005$&$\pm   0.02$&$\ldots$&&$\pm 1.1$&$\pm 0.005$&$\pm   0.02$&$\ldots$\nl
410&S0  & 0.870& 0.497&  20.70&   2.47&& 3.2& 0.370&  20.26&   2.48\nl
$\ldots$&$\ldots$&$\ldots$&$\pm 0.008$&$\pm   0.03$&$\ldots$&&$\pm 0.3$&$\pm 0.005$&$\pm   0.03$&$\ldots$\nl
412&E   & 0.843& 0.368&  20.22&   2.49&& 6.0& 0.795&  21.44&   2.45\nl
$\ldots$&$\ldots$&$\ldots$&$\pm 0.007$&$\pm   0.04$&$\ldots$&&$\pm 0.2$&$\pm 0.023$&$\pm   0.06$&$\ldots$\nl
433&S0/a& 0.851& 0.371&  19.95&   2.57&& 6.0& 0.907&  21.42&   2.51\nl
$\ldots$&$\ldots$&$\ldots$&$\pm 0.011$&$\pm   0.06$&$\ldots$&&$\pm 0.2$&$\pm 0.039$&$\pm   0.09$&$\ldots$\nl
440&S0/a& 0.843& 0.663&  21.67&   2.30&& 4.9& 0.843&  22.02&   2.30\nl
$\ldots$&$\ldots$&$\ldots$&$\pm 0.016$&$\pm   0.05$&$\ldots$&&$\pm 0.9$&$\pm 0.020$&$\pm   0.04$&$\ldots$\nl
454&S0/a& 0.807& 0.983&  21.16&   2.63&& 2.3& 0.613&  20.46&   2.64\nl
$\ldots$&$\ldots$&$\ldots$&$\pm 0.014$&$\pm   0.03$&$\ldots$&&$\pm 0.3$&$\pm 0.005$&$\pm   0.02$&$\ldots$\nl
463&S0  & 0.883& 0.667&  20.50&   2.66&& 3.2& 0.492&  20.03&   2.67\nl
$\ldots$&$\ldots$&$\ldots$&$\pm 0.007$&$\pm   0.02$&$\ldots$&&$\pm 0.3$&$\pm 0.004$&$\pm   0.02$&$\ldots$\nl
465&Sa  & 0.873& 0.581&  20.93&   2.47&& 4.2& 0.599&  20.98&   2.47\nl
$\ldots$&$\ldots$&$\ldots$&$\pm 0.009$&$\pm   0.03$&$\ldots$&&$\pm 0.3$&$\pm 0.009$&$\pm   0.03$&$\ldots$\nl
481&S0  & 0.810& 0.478&  21.06&   2.35&& 2.3& 0.334&  20.54&   2.35\nl
$\ldots$&$\ldots$&$\ldots$&$\pm 0.015$&$\pm   0.06$&$\ldots$&&$\pm 0.4$&$\pm 0.005$&$\pm   0.03$&$\ldots$\nl
493&E/S0& 0.809& 0.207&  20.09&   2.28&& 4.1& 0.214&  20.15&   2.27\nl
$\ldots$&$\ldots$&$\ldots$&$\pm 0.010$&$\pm   0.10$&$\ldots$&&$\pm 1.1$&$\pm 0.010$&$\pm   0.10$&$\ldots$\nl
523&S0/a& 0.905& 0.910&  21.42&   2.52&& 4.0& 0.863&  21.34&   2.52\nl
$\ldots$&$\ldots$&$\ldots$&$\pm 0.016$&$\pm   0.04$&$\ldots$&&$\pm 1.0$&$\pm 0.016$&$\pm   0.03$&$\ldots$\nl
531&E   & 0.887& 1.048&  20.63&   2.82&& 5.4& 1.682&  21.39&   2.79\nl
$\ldots$&$\ldots$&$\ldots$&$\pm 0.009$&$\pm   0.02$&$\ldots$&&$\pm 0.4$&$\pm 0.010$&$\pm   0.01$&$\ldots$\nl
534&E   & 0.888& 0.464&  20.77&   2.42&& 4.4& 0.486&  20.84&   2.42\nl
$\ldots$&$\ldots$&$\ldots$&$\pm 0.008$&$\pm   0.04$&$\ldots$&&$\pm 0.4$&$\pm 0.008$&$\pm   0.04$&$\ldots$\nl
536&E   & 0.874& 0.890&  20.63&   2.75&& 5.2& 1.355&  21.29&   2.73\nl
$\ldots$&$\ldots$&$\ldots$&$\pm 0.009$&$\pm   0.02$&$\ldots$&&$\pm 0.3$&$\pm 0.010$&$\pm   0.02$&$\ldots$\nl
549&Sab & 0.760& 1.592&  22.77&   2.35&& 3.5& 1.380&  22.54&   2.36\nl
$\ldots$&$\ldots$&$\ldots$&$\pm 0.058$&$\pm   0.07$&$\ldots$&&$\pm 0.5$&$\pm 0.033$&$\pm   0.06$&$\ldots$\nl
\enddata
\tablecomments{
Morphological types taken from Fabricant \etal\ (1999); restframe
$(B-V)_z$ colors taken from van Dokkum \etal\ (1998a); radii expressed
in arcsec and surface brightnesses in mag per square arcsec, except
for the fundamental plane parameters in which the radii have been
transformed to kpc and the surface brightnesses to $L_\odot/$pc$^2$;
these structural parameters were derived by fitting the integrated
profiles between $r=0\Sec 1$ and $r\approx 2\,r_{1/2}$.
}
\end{deluxetable}

\begin{deluxetable}{r l r r c r r r r r r r c c c c c}
\tablewidth{0pt}
\tiny
\tablecaption{1D Bulge-Disk Decompositions\label{bdstruct}}
\tablehead{
\colhead{}&
\colhead{}&
\multicolumn{3}{c}{Total}&
\colhead{}&
\multicolumn{2}{c}{Bulge}&
\colhead{}&
\multicolumn{2}{c}{Disk}&
\colhead{}&
\multicolumn{4}{c}{Bulge Fraction}\nl
\colhead{ID}&
\colhead{Type}&
\colhead{$r_{1/2}$}&
\colhead{$\langle \mu\rangle_{1/2}$}&
\colhead{$r_{1/2}\langle I\rangle_{1/2}^{0.76}$}&
\colhead{}&
\colhead{$r_{1/2}$}&
\colhead{$\langle \mu\rangle_{1/2}$}&
\colhead{}&
\colhead{$r_{1/2}$}&
\colhead{$\langle \mu\rangle_{1/2}$}&
\colhead{}&
\colhead{$\le0\Sec 2$}&
\colhead{$\le0\Sec 5$}&
\colhead{$\le2\,r_{1/2}$}&
\colhead{Total}&
\colhead{$\epsilon$}\nl
\colhead{(1)}&
\colhead{(2)}&
\colhead{(3)}&
\colhead{(4)}&
\colhead{(5)}&
\colhead{}&
\colhead{(6)}&
\colhead{(7)}&
\colhead{}&
\colhead{(8)}&
\colhead{(9)}&
\colhead{}&
\colhead{(10)}&
\colhead{(11)}&
\colhead{(12)}&
\colhead{(13)}&
\colhead{(14)}\nl}
\startdata
095&S0  & 0.704&  20.94&   2.55&& 0.986&  21.84&& 0.064&  17.75&& 0.46& 0.64& 0.77& 0.85& 0.15\nl
$\ldots$&&$\pm 0.056$&$\pm   0.48$&$\ldots$&&$\pm 0.078$&$\pm   0.15$&&$\pm 0.005$&$\pm   0.12$&&$\pm 0.04$&$\pm 0.02$&$\pm 0.01$&$\pm 0.01$&$\ldots$\nl
110&S0/a& 0.350&  19.45&   2.70&& 0.102&  18.36&& 0.419&  20.12&& 0.51& 0.31& 0.26& 0.23& 0.69\nl
$\ldots$&&$\pm 0.059$&$\pm   0.91$&$\ldots$&&$\pm 0.063$&$\pm   0.95$&&$\pm 0.014$&$\pm   0.16$&&$\pm 0.02$&$\pm 0.03$&$\pm 0.04$&$\pm 0.02$&$\ldots$\nl
129&S0/a& 0.489&  19.93&   2.70&& 0.726&  20.95&& 0.047&  16.68&& 0.48& 0.65& 0.73& 0.82& 0.27\nl
$\ldots$&&$\pm 0.035$&$\pm   0.42$&$\ldots$&&$\pm 0.059$&$\pm   0.14$&&$\pm 0.001$&$\pm   0.03$&&$\pm 0.02$&$\pm 0.01$&$\pm 0.01$&$\pm 0.01$&$\ldots$\nl
135&S0  & 0.418&  20.45&   2.47&& 0.490&  20.87&& 0.039&  18.01&& 0.76& 0.85& 0.88& 0.92& 0.43\nl
$\ldots$&&$\pm 0.046$&$\pm   0.69$&$\ldots$&&$\pm 0.048$&$\pm   0.20$&&$\pm 0.009$&$\pm   0.37$&&$\pm 0.05$&$\pm 0.03$&$\pm 0.02$&$\pm 0.01$&$\ldots$\nl
142&S0/a& 0.834&  21.40&   2.48&& 0.318&  20.47&& 1.094&  22.44&& 0.84& 0.64& 0.39& 0.34& 0.36\nl
$\ldots$&&$\pm 0.150$&$\pm   0.85$&$\ldots$&&$\pm 0.127$&$\pm   0.55$&&$\pm 0.072$&$\pm   0.25$&&$\pm 0.01$&$\pm 0.01$&$\pm 0.02$&$\pm 0.02$&$\ldots$\nl
164&S0/a& 2.251&  22.62&   2.54&& 2.541&  22.92&& 0.061&  17.77&& 0.46& 0.68& 0.91& 0.94& 0.12\nl
$\ldots$&&$\pm 0.188$&$\pm   0.45$&$\ldots$&&$\pm 0.207$&$\pm   0.13$&&$\pm 0.007$&$\pm   0.19$&&$\pm 0.06$&$\pm 0.03$&$\pm 0.00$&$\pm 0.01$&$\ldots$\nl
182&S0  & 0.355&  20.46&   2.40&& 0.392&  20.77&& 0.220&  21.91&& 0.87& 0.85& 0.86& 0.90& 0.43\nl
$\ldots$&&$\pm 0.024$&$\pm   0.34$&$\ldots$&&$\pm 0.023$&$\pm   0.10$&&$\pm 0.034$&$\pm   0.30$&&$\pm 0.02$&$\pm 0.02$&$\pm 0.02$&$\pm 0.02$&$\ldots$\nl
209&Sa  & 0.625&  20.65&   2.59&& 0.960&  22.06&& 0.427&  20.93&& 0.59& 0.50& 0.52& 0.64& 0.69\nl
$\ldots$&&$\pm 0.085$&$\pm   0.60$&$\ldots$&&$\pm 0.166$&$\pm   0.24$&&$\pm 0.030$&$\pm   0.11$&&$\pm 0.01$&$\pm 0.01$&$\pm 0.01$&$\pm 0.01$&$\ldots$\nl
211&S0  & 0.448&  20.12&   2.61&& 0.458&  20.41&& 0.432&  21.68&& 0.85& 0.77& 0.74& 0.78& 0.61\nl
$\ldots$&&$\pm 0.051$&$\pm   0.46$&$\ldots$&&$\pm 0.051$&$\pm   0.15$&&$\pm 0.052$&$\pm   0.18$&&$\pm 0.01$&$\pm 0.02$&$\pm 0.03$&$\pm 0.02$&$\ldots$\nl
212&E   & 0.683&  20.78&   2.59&& 0.781&  21.17&& 0.052&  18.03&& 0.70& 0.83& 0.89& 0.93& 0.09\nl
$\ldots$&&$\pm 0.036$&$\pm   0.32$&$\ldots$&&$\pm 0.034$&$\pm   0.09$&&$\pm 0.009$&$\pm   0.27$&&$\pm 0.04$&$\pm 0.02$&$\pm 0.01$&$\pm 0.01$&$\ldots$\nl
215&S0  & 0.457&  20.66&   2.45&& 0.700&  21.77&& 0.037&  16.95&& 0.47& 0.64& 0.71& 0.81& 0.29\nl
$\ldots$&&$\pm 0.036$&$\pm   0.46$&$\ldots$&&$\pm 0.056$&$\pm   0.15$&&$\pm 0.003$&$\pm   0.13$&&$\pm 0.03$&$\pm 0.01$&$\pm 0.01$&$\pm 0.01$&$\ldots$\nl
233&E/S0& 0.784&  20.40&   2.76&& 0.952&  20.94&& 0.113&  18.74&& 0.64& 0.76& 0.86& 0.90& 0.35\nl
$\ldots$&&$\pm 0.086$&$\pm   0.76$&$\ldots$&&$\pm 0.104$&$\pm   0.23$&&$\pm 0.014$&$\pm   0.17$&&$\pm 0.08$&$\pm 0.05$&$\pm 0.03$&$\pm 0.01$&$\ldots$\nl
234&Sb  & 0.996&  21.86&   2.42&& 0.035&  17.80&& 1.049&  22.03&& 0.55& 0.22& 0.06& 0.05& 0.40\nl
$\ldots$&&$\pm 0.113$&$\pm   0.51$&$\ldots$&&$\pm 0.041$&$\pm   2.22$&&$\pm 0.058$&$\pm   0.10$&&$\pm 0.01$&$\pm 0.01$&$\pm 0.01$&$\pm 0.02$&$\ldots$\nl
236&S0  & 0.776&  21.10&   2.55&& 0.944&  21.63&& 0.055&  17.87&& 0.59& 0.75& 0.86& 0.90& 0.29\nl
$\ldots$&&$\pm 0.047$&$\pm   0.37$&$\ldots$&&$\pm 0.055$&$\pm   0.11$&&$\pm 0.004$&$\pm   0.12$&&$\pm 0.04$&$\pm 0.02$&$\pm 0.01$&$\pm 0.01$&$\ldots$\nl
242&E   & 1.529&  21.92&   2.59&& 1.825&  22.42&& 0.263&  20.72&& 0.69& 0.70& 0.87& 0.91& 0.10\nl
$\ldots$&&$\pm 0.171$&$\pm   0.76$&$\ldots$&&$\pm 0.214$&$\pm   0.22$&&$\pm 0.015$&$\pm   0.24$&&$\pm 0.07$&$\pm 0.06$&$\pm 0.02$&$\pm 0.02$&$\ldots$\nl
256&E   & 1.380&  20.88&   2.86&& 1.551&  21.20&& 0.090&  18.00&& 0.63& 0.78& 0.91& 0.94& 0.14\nl
$\ldots$&&$\pm 0.073$&$\pm   0.33$&$\ldots$&&$\pm 0.071$&$\pm   0.09$&&$\pm 0.014$&$\pm   0.25$&&$\pm 0.05$&$\pm 0.03$&$\pm 0.01$&$\pm 0.01$&$\ldots$\nl
269&E/S0& 0.962&  20.39&   2.85&& 1.045&  20.62&& 0.093&  18.71&& 0.78& 0.87& 0.94& 0.96& 0.35\nl
$\ldots$&&$\pm 0.058$&$\pm   0.39$&$\ldots$&&$\pm 0.056$&$\pm   0.11$&&$\pm 0.019$&$\pm   0.28$&&$\pm 0.05$&$\pm 0.03$&$\pm 0.01$&$\pm 0.01$&$\ldots$\nl
292&S0/a& 0.486&  20.32&   2.58&& 0.297&  20.39&& 0.561&  21.10&& 0.64& 0.44& 0.36& 0.35& 0.26\nl
$\ldots$&&$\pm 0.137$&$\pm   1.35$&$\ldots$&&$\pm 0.218$&$\pm   0.93$&&$\pm 0.022$&$\pm   0.26$&&$\pm 0.02$&$\pm 0.03$&$\pm 0.04$&$\pm 0.01$&$\ldots$\nl
298&S0  & 0.554&  19.69&   2.83&& 0.835&  20.84&& 0.240&  19.53&& 0.61& 0.60& 0.68& 0.78& 0.67\nl
$\ldots$&&$\pm 0.052$&$\pm   0.56$&$\ldots$&&$\pm 0.088$&$\pm   0.19$&&$\pm 0.013$&$\pm   0.14$&&$\pm 0.03$&$\pm 0.02$&$\pm 0.01$&$\pm 0.01$&$\ldots$\nl
300&S0  & 0.453&  19.84&   2.69&& 0.584&  20.53&& 0.092&  18.66&& 0.66& 0.77& 0.82& 0.88& 0.41\nl
$\ldots$&&$\pm 0.028$&$\pm   0.43$&$\ldots$&&$\pm 0.039$&$\pm   0.13$&&$\pm 0.002$&$\pm   0.10$&&$\pm 0.03$&$\pm 0.02$&$\pm 0.01$&$\pm 0.01$&$\ldots$\nl
303&E   & 0.638&  20.59&   2.61&& 0.774&  21.11&& 0.091&  18.89&& 0.67& 0.78& 0.86& 0.90& 0.15\nl
$\ldots$&&$\pm 0.058$&$\pm   0.63$&$\ldots$&&$\pm 0.068$&$\pm   0.19$&&$\pm 0.012$&$\pm   0.16$&&$\pm 0.06$&$\pm 0.04$&$\pm 0.03$&$\pm 0.01$&$\ldots$\nl
309&E/S0& 0.502&  20.02&   2.68&& 0.540&  20.21&& 0.060&  18.91&& 0.87& 0.92& 0.94& 0.96& 0.14\nl
$\ldots$&&$\pm 0.023$&$\pm   0.31$&$\ldots$&&$\pm 0.020$&$\pm   0.08$&&$\pm 0.016$&$\pm   0.40$&&$\pm 0.04$&$\pm 0.02$&$\pm 0.02$&$\pm 0.01$&$\ldots$\nl
328&Sa  & 0.666&  20.72&   2.59&& 0.862&  21.53&& 0.392&  21.30&& 0.76& 0.69& 0.71& 0.80& 0.47\nl
$\ldots$&&$\pm 0.082$&$\pm   0.58$&$\ldots$&&$\pm 0.108$&$\pm   0.19$&&$\pm 0.046$&$\pm   0.20$&&$\pm 0.02$&$\pm 0.02$&$\pm 0.02$&$\pm 0.01$&$\ldots$\nl
335&S0/a& 0.539&  20.86&   2.46&& 0.541&  20.90&& 0.059&  20.61&& 0.94& 0.97& 0.98& 0.99& 0.37\nl
$\ldots$&&$\pm 0.040$&$\pm   0.51$&$\ldots$&&$\pm 0.033$&$\pm   0.14$&&$\pm 0.053$&$\pm   1.28$&&$\pm 0.06$&$\pm 0.04$&$\pm 0.02$&$\pm 0.01$&$\ldots$\nl
343&S0  & 0.404&  20.54&   2.43&& 0.492&  21.10&& 0.165&  20.93&& 0.77& 0.80& 0.83& 0.88& 0.44\nl
$\ldots$&&$\pm 0.038$&$\pm   0.54$&$\ldots$&&$\pm 0.045$&$\pm   0.16$&&$\pm 0.020$&$\pm   0.30$&&$\pm 0.03$&$\pm 0.02$&$\pm 0.02$&$\pm 0.02$&$\ldots$\nl
353&E/S0& 1.120&  21.05&   2.72&& 1.205&  21.24&& 0.134&  19.97&& 0.81& 0.88& 0.94& 0.96& 0.02\nl
$\ldots$&&$\pm 0.051$&$\pm   0.32$&$\ldots$&&$\pm 0.050$&$\pm   0.09$&&$\pm 0.021$&$\pm   0.22$&&$\pm 0.04$&$\pm 0.03$&$\pm 0.01$&$\pm 0.01$&$\ldots$\nl
356&Sa  & 0.604&  20.20&   2.71&& 0.392&  19.78&& 0.899&  22.07&& 0.90& 0.79& 0.65& 0.61& 0.46\nl
$\ldots$&&$\pm 0.044$&$\pm   0.31$&$\ldots$&&$\pm 0.035$&$\pm   0.11$&&$\pm 0.039$&$\pm   0.17$&&$\pm 0.00$&$\pm 0.00$&$\pm 0.01$&$\pm 0.02$&$\ldots$\nl
359&S0  & 0.574&  20.26&   2.67&& 0.769&  21.05&& 0.122&  19.05&& 0.61& 0.71& 0.79& 0.86& 0.20\nl
$\ldots$&&$\pm 0.056$&$\pm   0.71$&$\ldots$&&$\pm 0.082$&$\pm   0.22$&&$\pm 0.005$&$\pm   0.13$&&$\pm 0.06$&$\pm 0.04$&$\pm 0.03$&$\pm 0.01$&$\ldots$\nl
360&E   & 0.341&  20.22&   2.45&& 0.594&  21.66&& 0.050&  17.62&& 0.45& 0.61& 0.65& 0.77& 0.31\nl
$\ldots$&&$\pm 0.056$&$\pm   1.00$&$\ldots$&&$\pm 0.103$&$\pm   0.34$&&$\pm 0.007$&$\pm   0.21$&&$\pm 0.05$&$\pm 0.03$&$\pm 0.02$&$\pm 0.01$&$\ldots$\nl
366&S0/a& 0.630&  20.79&   2.55&& 0.718&  21.16&& 0.140&  20.46&& 0.80& 0.85& 0.90& 0.93& 0.28\nl
$\ldots$&&$\pm 0.038$&$\pm   0.44$&$\ldots$&&$\pm 0.044$&$\pm   0.12$&&$\pm 0.008$&$\pm   0.25$&&$\pm 0.04$&$\pm 0.03$&$\pm 0.02$&$\pm 0.01$&$\ldots$\nl
368&Sab & 0.542&  21.10&   2.39&& 0.627&  22.39&& 0.514&  21.56&& 0.53& 0.39& 0.34& 0.41& 0.63\nl
$\ldots$&&$\pm 0.139$&$\pm   0.97$&$\ldots$&&$\pm 0.317$&$\pm   0.62$&&$\pm 0.043$&$\pm   0.10$&&$\pm 0.01$&$\pm 0.01$&$\pm 0.02$&$\pm 0.02$&$\ldots$\nl
369&S0/a& 1.036&  21.86&   2.44&& 1.085&  21.98&& 0.079&  20.28&& 0.85& 0.92& 0.96& 0.98& 0.30\nl
$\ldots$&&$\pm 0.077$&$\pm   0.48$&$\ldots$&&$\pm 0.075$&$\pm   0.13$&&$\pm 0.024$&$\pm   0.41$&&$\pm 0.06$&$\pm 0.03$&$\pm 0.01$&$\pm 0.01$&$\ldots$\nl
371&Sa  & 0.956&  21.04&   2.66&& 0.468&  20.64&& 1.169&  21.92&& 0.81& 0.63& 0.37& 0.34& 0.54\nl
$\ldots$&&$\pm 0.118$&$\pm   0.61$&$\ldots$&&$\pm 0.147$&$\pm   0.42$&&$\pm 0.029$&$\pm   0.14$&&$\pm 0.00$&$\pm 0.01$&$\pm 0.02$&$\pm 0.01$&$\ldots$\nl
372&Sa  & 1.435&  21.92&   2.56&& 1.905&  22.72&& 0.593&  21.99&& 0.78& 0.71& 0.77& 0.84& 0.77\nl
$\ldots$&&$\pm 0.211$&$\pm   0.79$&$\ldots$&&$\pm 0.280$&$\pm   0.24$&&$\pm 0.088$&$\pm   0.32$&&$\pm 0.03$&$\pm 0.05$&$\pm 0.03$&$\pm 0.02$&$\ldots$\nl
375&E   & 4.979&  22.86&   2.82&& 5.267&  23.01&& 0.162&  19.22&& 0.54& 0.69& 0.96& 0.97& 0.05\nl
$\ldots$&&$\pm 0.313$&$\pm   0.37$&$\ldots$&&$\pm 0.295$&$\pm   0.10$&&$\pm 0.046$&$\pm   0.47$&&$\pm 0.13$&$\pm 0.07$&$\pm 0.01$&$\pm 0.01$&$\ldots$\nl
381&E/S0& 0.512&  20.34&   2.60&& 0.550&  20.50&& 0.088&  20.02&& 0.88& 0.92& 0.94& 0.96& 0.37\nl
$\ldots$&&$\pm 0.063$&$\pm   0.89$&$\ldots$&&$\pm 0.060$&$\pm   0.25$&&$\pm 0.041$&$\pm   0.64$&&$\pm 0.10$&$\pm 0.07$&$\pm 0.05$&$\pm 0.02$&$\ldots$\nl
397&S0/a& 0.409&  20.50&   2.45&& 0.111&  18.96&& 0.540&  21.45&& 0.68& 0.44& 0.35& 0.29& 0.57\nl
$\ldots$&&$\pm 0.035$&$\pm   0.45$&$\ldots$&&$\pm 0.026$&$\pm   0.36$&&$\pm 0.014$&$\pm   0.10$&&$\pm 0.01$&$\pm 0.01$&$\pm 0.01$&$\pm 0.01$&$\ldots$\nl
408&S0  & 0.382&  19.82&   2.63&& 0.523&  20.66&& 0.096&  18.86&& 0.63& 0.74& 0.77& 0.85& 0.54\nl
$\ldots$&&$\pm 0.020$&$\pm   0.37$&$\ldots$&&$\pm 0.030$&$\pm   0.12$&&$\pm 0.002$&$\pm   0.08$&&$\pm 0.02$&$\pm 0.01$&$\pm 0.01$&$\pm 0.01$&$\ldots$\nl
409&E   & 0.498&  21.27&   2.30&& 0.542&  21.49&& 0.223&  22.60&& 0.90& 0.90& 0.92& 0.94& 0.10\nl
$\ldots$&&$\pm 0.028$&$\pm   0.41$&$\ldots$&&$\pm 0.026$&$\pm   0.11$&&$\pm 0.041$&$\pm   0.62$&&$\pm 0.04$&$\pm 0.03$&$\pm 0.02$&$\pm 0.02$&$\ldots$\nl
410&S0  & 0.488&  20.67&   2.47&& 0.575&  21.13&& 0.157&  20.84&& 0.80& 0.83& 0.87& 0.91& 0.43\nl
$\ldots$&&$\pm 0.032$&$\pm   0.46$&$\ldots$&&$\pm 0.037$&$\pm   0.13$&&$\pm 0.013$&$\pm   0.29$&&$\pm 0.04$&$\pm 0.03$&$\pm 0.02$&$\pm 0.02$&$\ldots$\nl
412&E   & 0.767&  21.39&   2.45&& 1.091&  22.32&& 0.056&  17.66&& 0.42& 0.61& 0.76& 0.84& 0.32\nl
$\ldots$&&$\pm 0.072$&$\pm   0.56$&$\ldots$&&$\pm 0.096$&$\pm   0.17$&&$\pm 0.007$&$\pm   0.22$&&$\pm 0.06$&$\pm 0.03$&$\pm 0.01$&$\pm 0.01$&$\ldots$\nl
433&S0/a& 1.021&  21.61&   2.51&& 1.397&  22.46&& 0.072&  17.95&& 0.40& 0.59& 0.78& 0.85& 0.11\nl
$\ldots$&&$\pm 0.189$&$\pm   1.10$&$\ldots$&&$\pm 0.251$&$\pm   0.34$&&$\pm 0.016$&$\pm   0.36$&&$\pm 0.11$&$\pm 0.06$&$\pm 0.03$&$\pm 0.01$&$\ldots$\nl
440&S0/a& 1.108&  22.45&   2.29&& 1.328&  22.98&& 0.061&  18.80&& 0.54& 0.72& 0.87& 0.91& 0.34\nl
$\ldots$&&$\pm 0.109$&$\pm   0.58$&$\ldots$&&$\pm 0.128$&$\pm   0.17$&&$\pm 0.007$&$\pm   0.17$&&$\pm 0.06$&$\pm 0.03$&$\pm 0.01$&$\pm 0.01$&$\ldots$\nl
454&S0/a& 0.595&  20.42&   2.64&& 0.990&  22.27&& 0.461&  20.61&& 0.47& 0.37& 0.37& 0.50& 0.58\nl
$\ldots$&&$\pm 0.068$&$\pm   0.50$&$\ldots$&&$\pm 0.182$&$\pm   0.26$&&$\pm 0.020$&$\pm   0.07$&&$\pm 0.00$&$\pm 0.00$&$\pm 0.00$&$\pm 0.01$&$\ldots$\nl
463&S0  & 0.639&  20.43&   2.66&& 0.714&  20.74&& 0.229&  21.18&& 0.87& 0.87& 0.91& 0.94& 0.48\nl
$\ldots$&&$\pm 0.044$&$\pm   0.46$&$\ldots$&&$\pm 0.046$&$\pm   0.13$&&$\pm 0.033$&$\pm   0.46$&&$\pm 0.04$&$\pm 0.03$&$\pm 0.02$&$\pm 0.02$&$\ldots$\nl
465&Sa  & 0.609&  21.01&   2.47&& 0.660&  21.22&& 0.133&  21.10&& 0.87& 0.90& 0.94& 0.96& 0.48\nl
$\ldots$&&$\pm 0.050$&$\pm   0.64$&$\ldots$&&$\pm 0.053$&$\pm   0.18$&&$\pm 0.019$&$\pm   0.59$&&$\pm 0.07$&$\pm 0.05$&$\pm 0.03$&$\pm 0.02$&$\ldots$\nl
481&S0  & 0.260&  20.25&   2.33&& 0.157&  20.76&& 0.280&  20.65&& 0.32& 0.22& 0.22& 0.22& 0.74\nl
$\ldots$&&$\pm 0.036$&$\pm   0.65$&$\ldots$&&$\pm 0.090$&$\pm   0.70$&&$\pm 0.004$&$\pm   0.12$&&$\pm 0.03$&$\pm 0.04$&$\pm 0.04$&$\pm 0.02$&$\ldots$\nl
493&E/S0& 0.193&  19.98&   2.28&& 0.374&  21.66&& 0.059&  18.74&& 0.48& 0.61& 0.58& 0.73& 0.44\nl
$\ldots$&&$\pm 0.049$&$\pm   1.85$&$\ldots$&&$\pm 0.119$&$\pm   0.68$&&$\pm 0.005$&$\pm   0.17$&&$\pm 0.08$&$\pm 0.05$&$\pm 0.05$&$\pm 0.01$&$\ldots$\nl
523&S0/a& 1.044&  21.64&   2.51&& 1.171&  21.96&& 0.197&  21.08&& 0.80& 0.83& 0.91& 0.94& 0.25\nl
$\ldots$&&$\pm 0.079$&$\pm   0.52$&$\ldots$&&$\pm 0.089$&$\pm   0.15$&&$\pm 0.013$&$\pm   0.30$&&$\pm 0.05$&$\pm 0.04$&$\pm 0.02$&$\pm 0.02$&$\ldots$\nl
531&E   & 1.549&  21.26&   2.80&& 1.830&  21.71&& 0.182&  19.30&& 0.61& 0.69& 0.88& 0.92& 0.25\nl
$\ldots$&&$\pm 0.088$&$\pm   0.39$&$\ldots$&&$\pm 0.102$&$\pm   0.11$&&$\pm 0.013$&$\pm   0.10$&&$\pm 0.04$&$\pm 0.04$&$\pm 0.01$&$\pm 0.01$&$\ldots$\nl
534&E   & 0.620&  21.21&   2.41&& 0.878&  22.17&& 0.032&  16.68&& 0.44& 0.63& 0.74& 0.83& 0.24\nl
$\ldots$&&$\pm 0.043$&$\pm   0.38$&$\ldots$&&$\pm 0.060$&$\pm   0.12$&&$\pm 0.003$&$\pm   0.15$&&$\pm 0.02$&$\pm 0.01$&$\pm 0.01$&$\pm 0.01$&$\ldots$\nl
536&E   & 1.266&  21.18&   2.73&& 1.433&  21.52&& 0.096&  18.57&& 0.64& 0.78& 0.91& 0.94& 0.27\nl
$\ldots$&&$\pm 0.063$&$\pm   0.31$&$\ldots$&&$\pm 0.064$&$\pm   0.09$&&$\pm 0.012$&$\pm   0.19$&&$\pm 0.04$&$\pm 0.03$&$\pm 0.01$&$\pm 0.01$&$\ldots$\nl
549&Sab & 0.866&  21.92&   2.34&& 0.142&  19.90&& 1.028&  22.50&& 0.74& 0.47& 0.21& 0.17& 0.16\nl
$\ldots$&&$\pm 0.120$&$\pm   0.70$&$\ldots$&&$\pm 0.080$&$\pm   0.92$&&$\pm 0.049$&$\pm   0.15$&&$\pm 0.01$&$\pm 0.02$&$\pm 0.03$&$\pm 0.01$&$\ldots$\nl
\enddata
\tablecomments{
Morphological types taken from Fabricant \etal\ (1999); restframe
$(B-V)_z$ colors taken from van Dokkum \etal\ (1998a); radii expressed
in arcsec and surface brightnesses in mag per square arcsec, except
for the fundamental plane parameters in which the radii have been
transformed to kpc and the surface brightnesses to $L_\odot/$pc$^2$;
these structural parameters were derived by fitting the integrated
profiles between $r=0\Sec 1$ and $r\approx 2\,r_{1/2}$. Apparent
ellipticities derived from direct image fitting and are
luminosity-weighted mean values for $\epsilon$.
}
\end{deluxetable}

\clearpage

\begin{figure}
\caption[Color images of the CL1358+62 FP galaxy sample]{
Color images of the fundamental plane sample of CL1358+62 galaxies
generated from the F606W and F814W HST imaging. The boxes are $15''$
on each side. The galaxies classified as E+A are ID \# 209, 328, and
343. Galaxy \# 234 is a star-forming emission line galaxy. Galaxy \#
375 is the BCG.
\label{images}}
\end{figure}
\clearpage

\clearpage

\clearpage

\begin{center}
\begin{minipage}{\textwidth}
\vspace{1.3in}
\mkfigbox{7in}{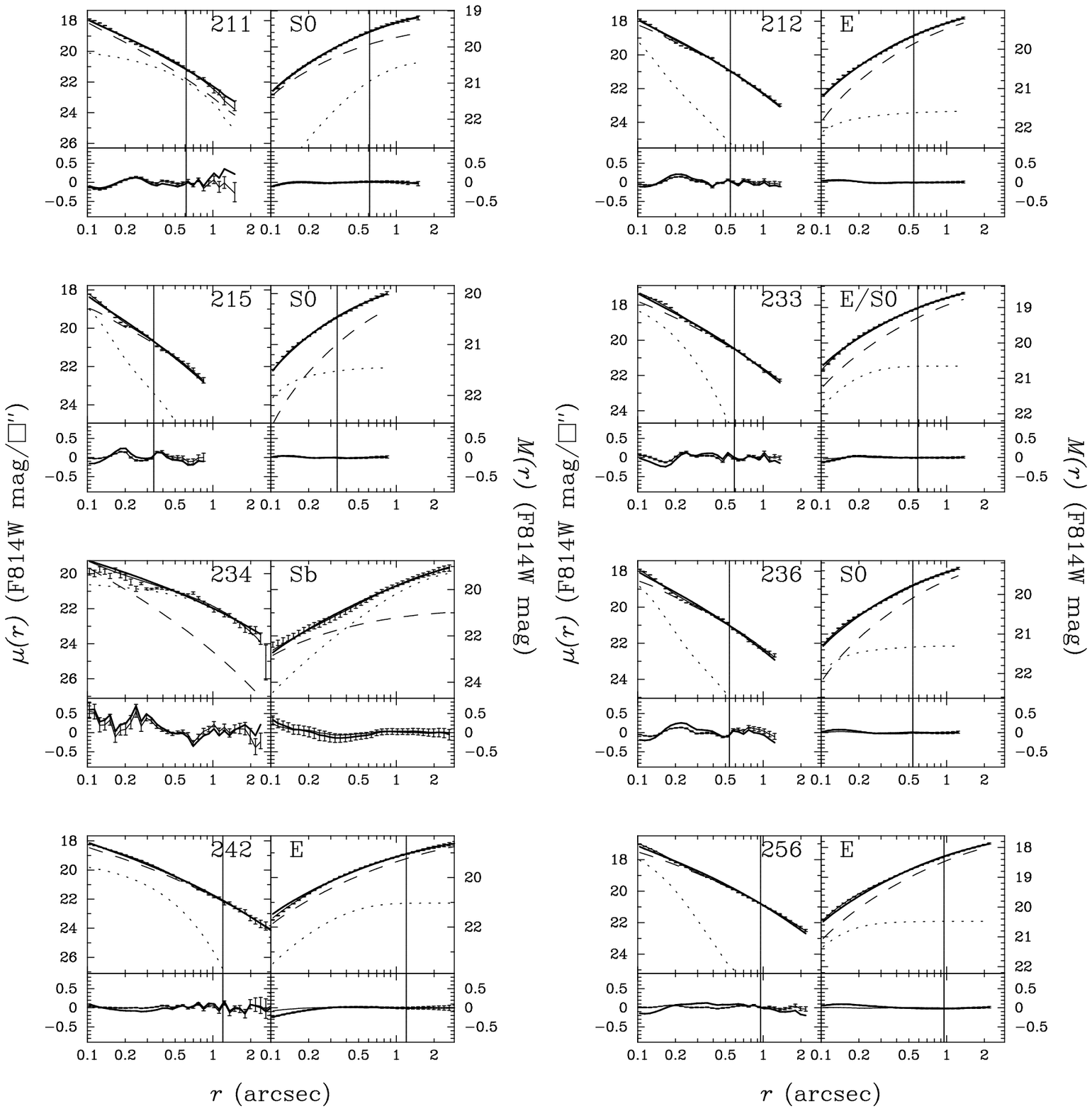}
\vspace{0.15in}
\Fig
\end{minipage}
\end{center}
\clearpage

\begin{center}
\begin{minipage}{\textwidth}
\vspace{1.3in}
\mkfigbox{7in}{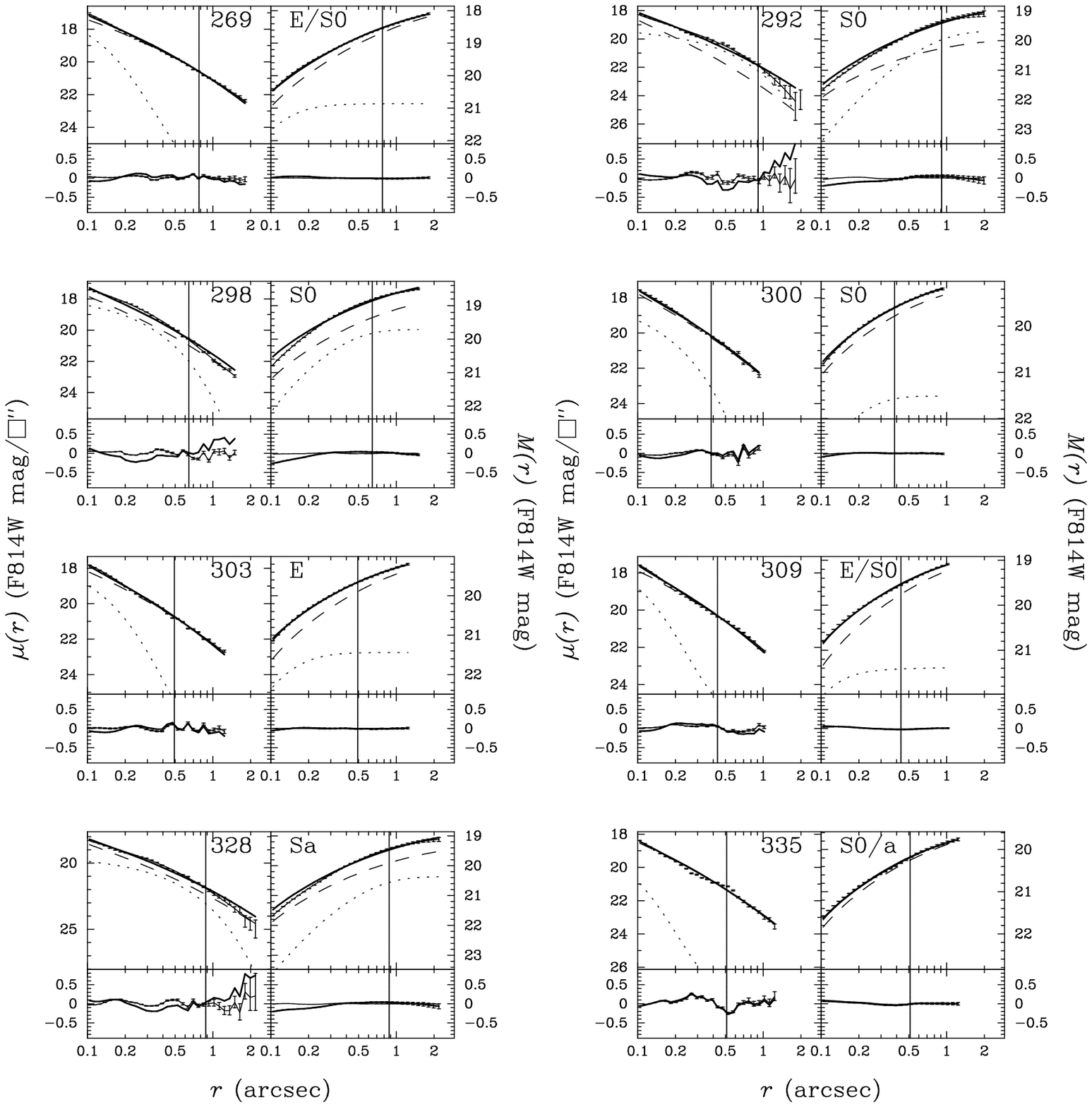}
\vspace{0.15in}
\Fig
\end{minipage}
\end{center}
\clearpage

\begin{center}
\begin{minipage}{\textwidth}
\vspace{1.3in}
\mkfigbox{7in}{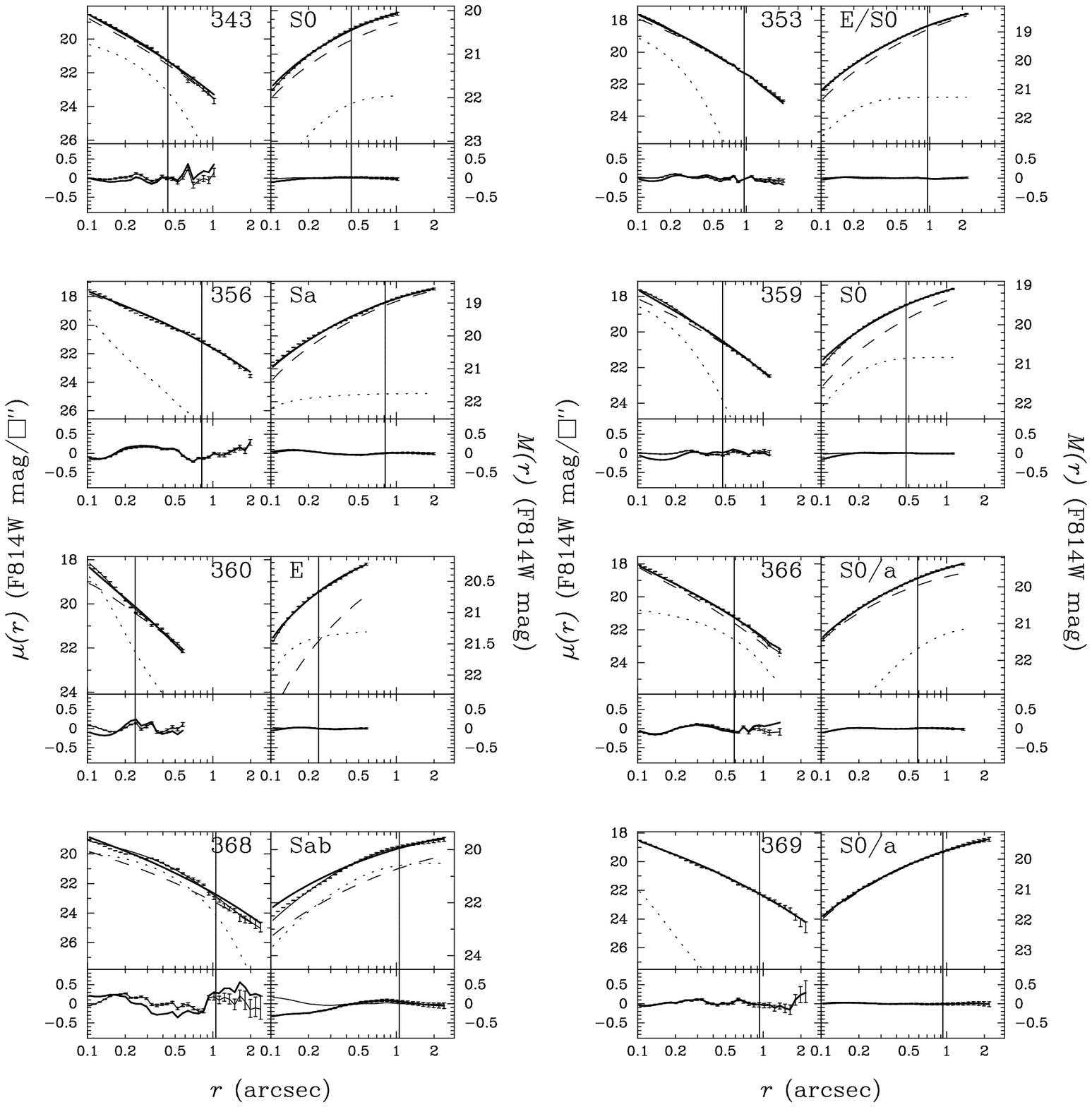}
\vspace{0.15in}
\Fig
\end{minipage}
\end{center}
\clearpage

\begin{center}
\begin{minipage}{\textwidth}
\vspace{1.3in}
\mkfigbox{7in}{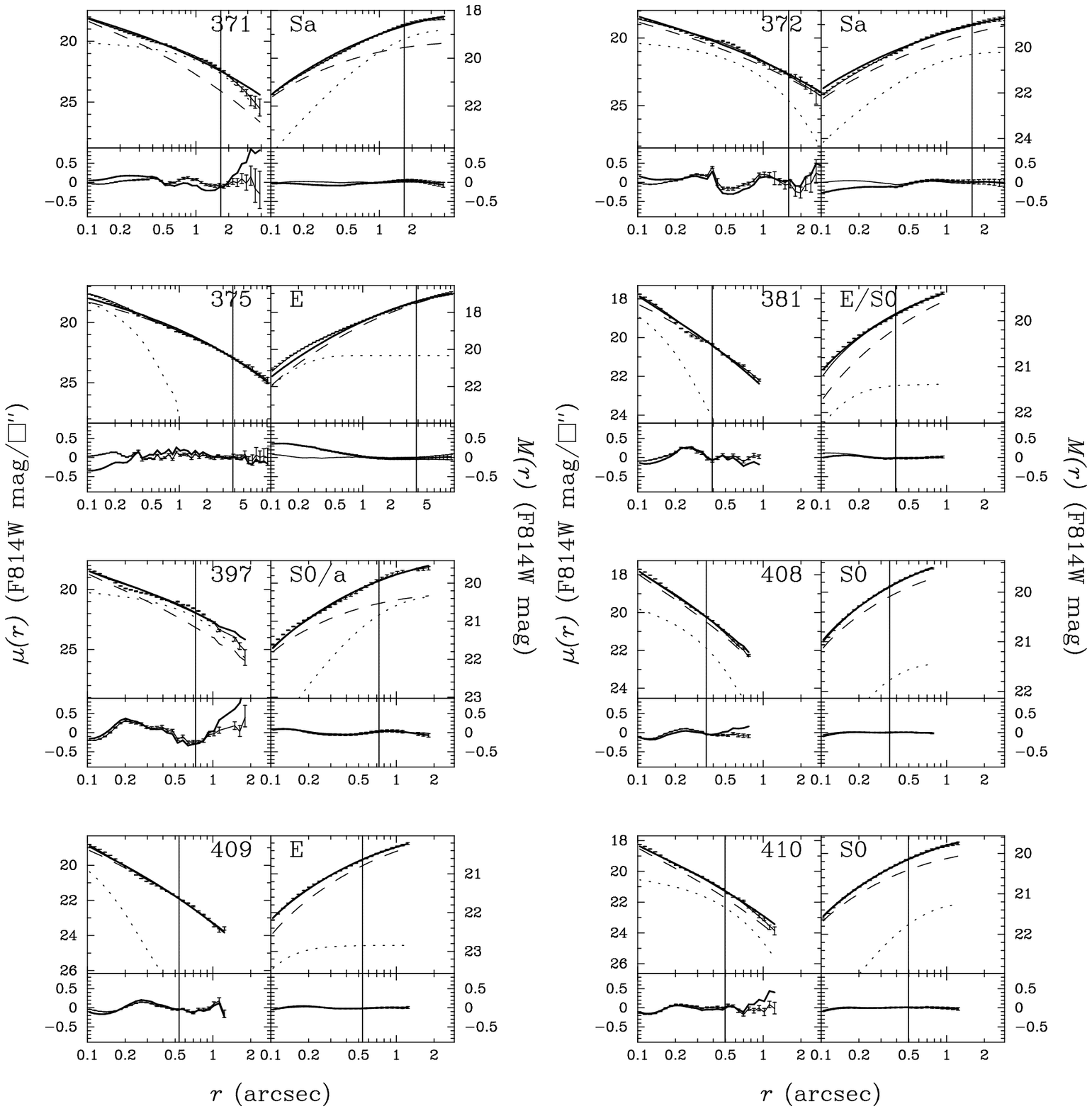}
\vspace{0.15in}
\Fig
\end{minipage}
\end{center}
\clearpage

\begin{center}
\begin{minipage}{\textwidth}
\vspace{1.3in}
\mkfigbox{7in}{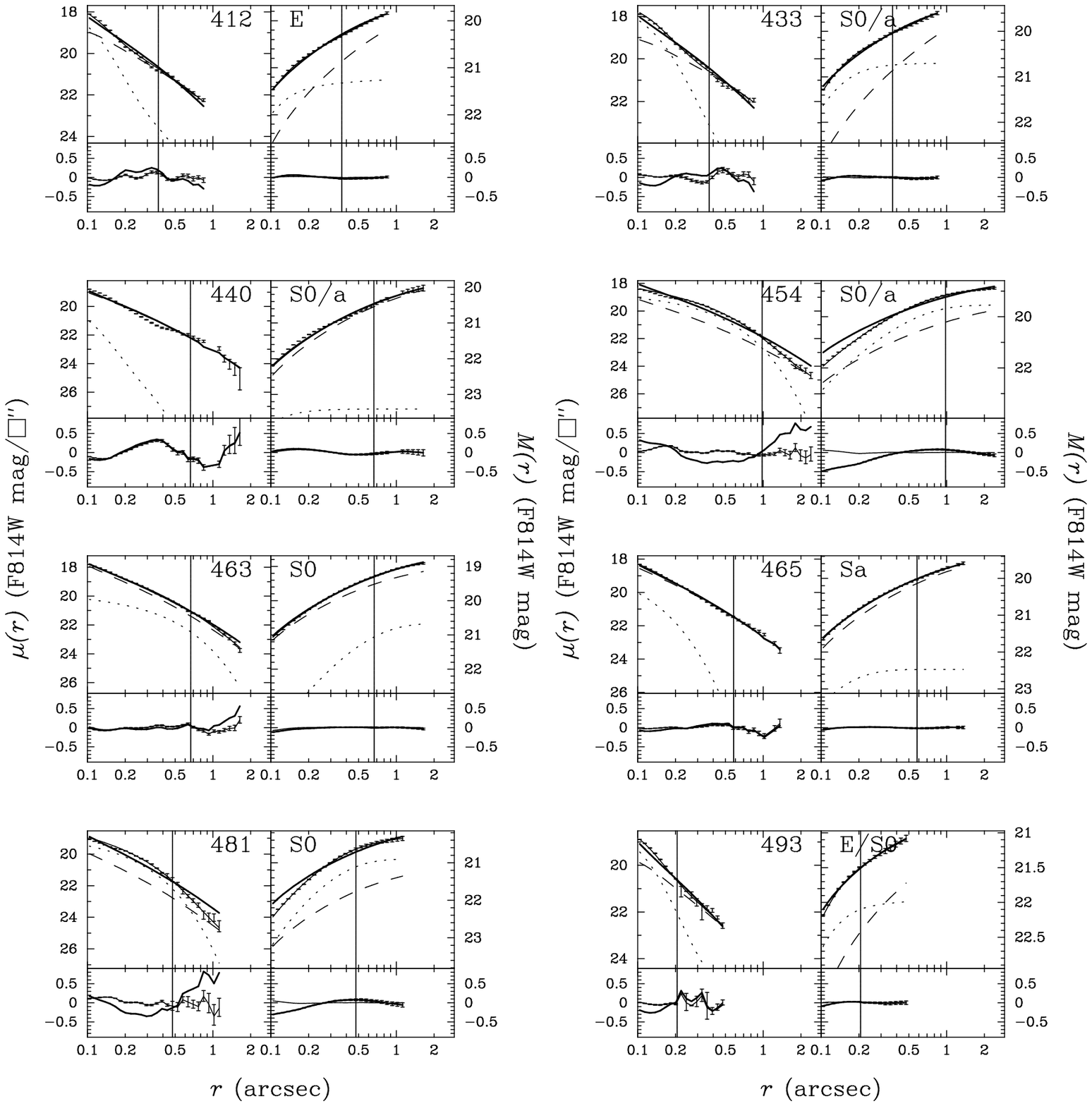}
\vspace{0.15in}
\Fig
\end{minipage}
\end{center}
\clearpage

\begin{center}
\begin{minipage}{\textwidth}
\vspace{1.3in}
\mkfigbox{7in}{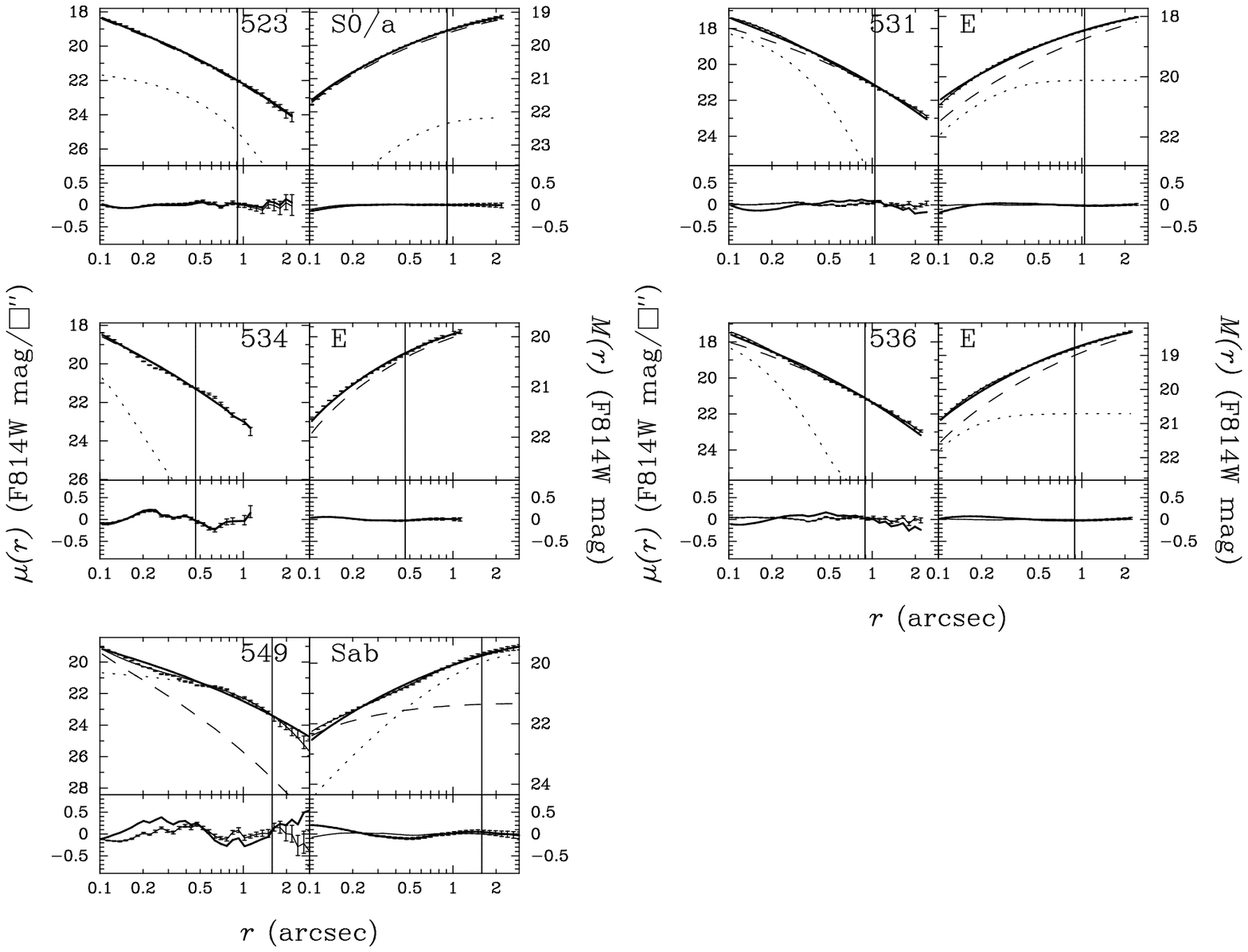}
\vspace{0.15in}
\Fig
\end{minipage}
\end{center}
\clearpage

\begin{figure}
\caption[F814W residual maps of the CL1358+62 FP galaxies]{
F814W residuals from the $r^{1/4}$-law image modeling. Note the disky
residuals in many galaxies. The boxes are $15''$ on each side. The
galaxies classified as E+A are ID \# 209, 328, and 343. Galaxy \# 234
is a star-forming emission line galaxy. Galaxy \# 375 is the BCG.
\label{residuals}}
\end{figure}
\clearpage

\clearpage

\clearpage

\clearpage

\end{document}